\pdfoutput=1
\documentclass[aps,prd,amsmath,floats,floatfix, twocolumn,
superscriptaddress,nofootinbib,showpacs,longbibliography]{revtex4-2}

\usepackage[T1]{fontenc}
\usepackage[utf8]{inputenc}
\usepackage{lmodern}

\usepackage{verbatim}

\usepackage[dvipsnames, usenames]{xcolor}
\definecolor{linkcolor}{rgb}{0.0,0.3,0.5}
\usepackage[hypertexnames=false, unicode, colorlinks=true, linkcolor=linkcolor,
citecolor=linkcolor, filecolor=linkcolor,urlcolor=linkcolor,
pdfusetitle]{hyperref}

\usepackage[all]{hypcap}
\usepackage{graphicx}
\usepackage{xspace}
\usepackage{amssymb}
\usepackage[normalem]{ulem} %for \sout
\usepackage{bm} % boldmath

\usepackage{microtype}

\usepackage[english]{babel}
\usepackage{blindtext}
\usepackage{acronym}
\usepackage[caption=false]{subfig}
\usepackage{float}

\graphicspath{%
  {figures/}%
}

\DeclareMathAlphabet{\mathpzc}{OT1}{pzc}{m}{it}

\newcommand{\chieff}{\chi_{\mathrm{eff}}}

\begin{document}

\title{Incorporating waveform calibration error in gravitational-wave modeling and inference for SEOBNRv4}

\newcommand{\URI}{\affiliation{Department of Physics and
    Center for Computational Research, East Hall,
    University of Rhode Island, Kingston, RI 02881}}

\newcommand{\AEI}{\affiliation{Max Planck Institute for Gravitational Physics
    (Albert Einstein Institute), Am M\"uhlenberg 1, Potsdam 14476, Germany}}

\author{Ritesh Bachhar}
\email{riteshbachhar@uri.edu}
\URI

\author{Michael P\"{u}rrer}
\URI
\AEI

\author{Stephen R. Green}
\affiliation{School of Mathematical Sciences, University of Nottingham, University Park, Nottingham NG7 2RD, United Kingdom}
\AEI

% Because hyperref only gets the *last* author, we need to be explicit.
\hypersetup{pdfauthor={Bachhar et al.}}

\date{\today}

%==========================================================================
\begin{abstract}
As \ac{GW} detector networks continue to improve in sensitivity, the demand on the accuracy of waveform models which predict the \ac{GW} signals from compact binary coalescences is becoming more stringent.
At high \acp{SNR} discrepancies between waveform models and the true solutions of Einstein's equations can introduce significant systematic biases in \ac{PE}. 
These biases affect the inferred astrophysical properties, including matter effects, and can also lead to erroneous claims of deviations from general relativity, impacting the interpretation of astrophysical populations and cosmological parameters. 
While efforts to address these biases have focused on developing more precise models, we explore an alternative strategy to account for uncertainties in waveform models, particularly from calibrating an \ac{EOB} model against \ac{NR} data.
We introduce an efficient method for modeling and marginalizing over waveform uncertainty in the \texttt{SEOBNRv4} model, which captures the dominant $(2,2)$ mode for non-precessing quasi-circular \acp{BBH}. Our approach uses \ac{GPR} to model amplitude and phase deviations in the Fourier domain. 
This method mitigates systematic biases in \ac{PE} and increases posterior variance by incorporating a broader distribution of waveforms, consistent with previous findings. 
This study emphasizes the importance of incorporating waveform uncertainties in GW data analysis and presents a novel, practical framework to include these uncertainties in Bayesian PE for EOB models, with broad applicability.
\end{abstract}

\maketitle

% ======================
%  ACRONYMS
% ======================
\acrodef{LSC}[LSC]{LIGO Scientific Collaboration}
\acrodef{aLIGO}{Advanced Laser Interferometer Gravitational wave Observatory}
\acrodef{aVirgo}{Advanced Virgo}
\acrodef{LIGO}[LIGO]{Laser Interferometer Gravitational-Wave Observatory}
\acrodef{IFO}[IFO]{interferometer}
\acrodef{LHO}[LHO]{LIGO-Hanford}
\acrodef{LLO}[LLO]{LIGO-Livingston}
\acrodef{O2}[O2]{second observing run}
\acrodef{O1}[O1]{first observing run}
\acrodef{BH}[BH]{black hole}
\acrodef{BBH}[BBH]{binary black hole}
\acrodef{BNS}[BNS]{binary neutron star}
\acrodef{NS}[NS]{neutron star}
\acrodef{BHNS}[BHNS]{black hole--neutron star binaries}
\acrodef{NSBH}[NSBH]{neutron star--black hole binary}
\acrodef{PBH}[PBH]{primordial black hole binaries}
\acrodef{CBC}[CBC]{compact binary coalescence}
\acrodef{GW}[GW]{gravitational wave}
\acrodef{CWB}[cWB]{coherent WaveBurst}
\acrodef{SNR}[SNR]{signal-to-noise ratio}
\acrodef{FAR}[FAR]{false alarm rate}
\acrodef{IFAR}[IFAR]{inverse false alarm rate}
\acrodef{FAP}[FAP]{false alarm probability}
\acrodef{PSD}[PSD]{power spectral density}
\acrodef{ASD}[ASD]{amplitude spectral density}
\acrodef{GR}[GR]{general relativity}
\acrodef{NR}[NR]{numerical relativity}
\acrodef{PN}[PN]{post-Newtonian}
\acrodef{EOB}[EOB]{effective-one-body}
\acrodef{ROM}[ROM]{reduced-order-model}
\acrodef{IMR}[IMR]{inspiral-merger-ringdown}
\acrodef{PDF}[PDF]{probability density function}
\acrodef{PE}[PE]{parameter estimation}
\acrodef{CL}[CL]{credible level}
\acrodef{EOS}[EOS]{equation of state}
\acrodef{LAL}[LAL]{LSC Algorithm Library}
\acrodef{ET}[ET]{Einstein Telescope}
\acrodef{CE}[CE]{Cosmic Explorer}
\acrodef{MAP}[MAP]{maximum a posteriori}
\acrodef{KDE}[KDE]{kernel density estimate}
\acrodef{CDF}[CDF]{cumulative distribution function}
\acrodef{CCE}[CCE]{Cauchy characteristic extraction}
\acrodef{GPR}[GPR]{Gaussian process regression}
\acrodef{GP}[GP]{Gaussian process}
\acrodef{PN-NR}[PN-NR]{post-Newtonian - numerical relativity}

%==========================================================================
\section{Introduction}
\label{sec:introduction}

To date approximately 100 \ac{GW} signals from the coalescence of compact binaries have been observed by the LIGO-Virgo-KAGRA (LVK) collaboration~\cite{LIGOScientific:2014pky,VIRGO:2014yos,KAGRA:2020tym,KAGRA:2021vkt,LIGOScientific:2021usb}.
Accurate waveform models are essential for robust estimation of source parameters~\cite{Chatziioannou:2024hju,Ashton:2018jfp,bilby_pipe_paper,Thrane:2018qnx,Veitch:2014wba}. 
However, these models are in practice subject to systematic uncertainties.
Inspiral-merger-ringdown waveform models are in general constructed from an approximate ansatz which is tuned to a finite set of \ac{NR} simulation data~\cite{Chatziioannou:2024hju}. Internally, approximate fits against \ac{NR} data are used to predict model coefficients over the binary parameter space. We refer to errors arising at each \ac{NR} point and in the fits over parameter space as \emph{waveform calibration errors}.
This is the case for the phenomenological~\cite{Khan:2015jqa,Pratten:2020fqn,London:2017bcn,Garcia-Quiros:2020qpx,Hannam:2013oca,Pratten:2020ceb,Hamilton:2021pkf,Estelles:2020osj,Estelles:2020twz,Estelles:2021gvs,Dietrich:2017aum,Dietrich:2019kaq,Thompson:2020nei} and \ac{EOB} models~\cite{Bohe:2016gbl,Cotesta:2018fcv,Pompili:2023tna,Ossokine:2020kjp,Ramos-Buades:2023ehm,Steinhoff:2016rfi,Matas:2020wab,Nagar:2018zoe,Nagar:2018plt,Akcay:2018yyh,Akcay:2020qrj,Gamba:2021ydi} in use today. 
In contrast, \ac{NR} surrogates model a set of \ac{NR} simulation data directly in a data driven way, with imperfections arising from errors in reduced bases and fits over parameter space and have been shown to be highly accurate~\cite{Field:2013cfa,Blackman:2017dfb,Blackman:2017pcm,Varma:2019csw,Varma:2018mmi,Islam:2021mha}.

\ac{GW} catalogs have so far used techniques where posterior samples for different \ac{GW} models are combined with equal weights~\cite{KAGRA:2021vkt}, weighted by their respective evidences~\cite{Ashton:2019leq}, or by averaging the likelihood~\cite{Jan:2020bdz}. Alternatively, one can sample of a set of models in a single joint Bayesian analysis~\cite{Ashton:2021cub,Hoy:2022tst}, and prioritize the most accurate model in each region of parameter space~\cite{Hoy:2024vpc}.
In this study, we incorporate waveform calibration uncertainties into Bayesian inference for the \texttt{SEOBNRv4} model, focusing on improving the accuracy (and reduce bias) of parameter estimates at the expense of precision (expecting wider posterior distributions).

As \ac{GW} detectors are becoming more sensitive the rate of events is ever increasing, as well as the \ac{SNR} for golden events. Waveform systematics arise when there are discrepancies between a particular waveform model and the true \ac{GW} signal from solving Einstein's equations, assuming that \ac{GR} is the correct theory of gravity. These errors can significantly impact the posterior distributions of the inferred source parameters, leading to biased results~\cite{LIGOScientific:2016ebw,Purrer:2019jcp,Dhani:2024jja,Kapil:2024zdn,Ferguson:2020xnm,Hu:2022rjq,Jan:2023raq}. The motivation for this work is rooted in the need to refine posterior distributions of approximate waveform models for moderate to high \ac{SNR} events where waveform systematics will arise.
One way to reduce biases is to increase the accuracy of waveform models. This requires improvements in approximation theory, waveform construction, and most importantly, an extensive set of \ac{NR} simulations covering the the high dimensional binary parameter space, a requirement which is strongly computationally limited. Here we follow an alternative approach by including uncertainties inherent in an existing model itself.
We adopt a novel approach using Gaussian process regression to interpolate amplitude and phase differences between the calibrated \texttt{SEOBNRv4} model and an uncalibrated version of this \ac{EOB} model where calibration parameters are sampled from calibration posterior distributions at points in parameter space where \ac{NR} waveforms are available. A calibration posterior arises from a likelihood function which encodes discrepancies between the uncalibrated \ac{EOB} model and an \ac{NR} simulation. This approach enables the transition from a deterministic to a probabilistic waveform model, effectively capturing the model uncertainty.

The idea of marginalizing over waveform uncertainty in parameter estimation was first proposed by Moore \& Gair~\cite{Moore:2014pda,Moore:2015sza} in an idealized setting. Their approach is elegant and relies on
the approximation of the strain difference by a Gaussian process which allows for analytical marginalization
in a likelihood function which is marginalized over waveform uncertainty.
For the sake of efficiency, and because strain differences are difficult to model directly, we instead apply a frequency-dependent amplitude and phase error to each template waveform which enters the \ac{GW} likelihood function. This methodology is inspired by established techniques for marginalizing over detector calibration uncertainty~\cite{SplineCalMarg-T1400682,Vitale:2011wu} and applications of this parametrization. In particular, \citet{Edelman:2020aqj} augmented an existing waveform model with spline parametrized amplitude and phase deviations in the Fourier domain to constrain unmodeled physics, while \citet{Read:2023hkv} used a similar ansatz to quantify the amplitude and phase accuracy of waveform models for neutron star inspirals.
In a similar spirit~\citet{Owen:2023mid} employed a \ac{PN}-based method to mitigate waveform systematics has recently been introduced by where the authors propose to marginalize over higher order \ac{PN} coefficients.

Several studies have investigated the use of probabilistic waveform models. For instance, a \ac{GPR} model based on spin-aligned phenomenological waveforms was built by~\cite{Doctor:2017csx} along with a greedy method for predicting where new training waveforms should be placed to reduce modeling error. 
Additionally, \cite{Williams:2019vub} trained a model on precessing \ac{NR} simulations using a \ac{GP} which is joint in model parameters and time. A combination of active learning with \ac{GPR} was explored in~\cite{Andrade:2023sal} to train an \ac{EOB} model against \ac{NR} data. More recently, \ac{GPR} was employed to developing a probabilistic extension to the phenomenological modeling workflow for non-spinning \acp{BH} while taking into account the estimated numerical error in \ac{NR} simulations.

% forward citation to LP paper -- in prep
This study is similar to the one discussed in~\citet{Pompili1024inprep}. Both studies introduce a method for modeling waveform uncertainty arising in the calibration of \ac{EOB} models against \ac{NR} data and use these models in a Bayesian inference study. While the overall goals align strongly, there are significant differences in how the uncertainty models are constructed and how uncertainty is incorporated into \ac{PE}.

The stochasticity in our uncertainty model arises from a multivariate normal approximation of the distribution of amplitude and phase differences at points in parameter space where \ac{NR} simulations were available at the time of construction of the \texttt{SEOBNRv4} model. Subsequently, we use \ac{GPR} to interpolate the means and covariance matrices of amplitude and phase deviations over parameter space.

By integrating waveform calibration errors into the \texttt{SEOBNRv4} model we explore additional variations in the waveform templates, with the aim of improving the accuracy of \ac{GW} parameter estimates at the expense of precision. While our method is presented for a particular \ac{EOB} model, it is broadly applicable for any waveform model where a distribution of waveforms around the best fit waveform is available at calibration points over the binary parameter space. This approach not only provides a comprehensive uncertainty quantification but also paves the way for more robust tests of general relativity and enhanced astrophysical inferences.

This paper is organized as follows. In Sec.~\ref{sec:methods} we present a comprehensive summary of methods
pertaining to the comparison of waveforms, we discuss \ac{GPR} and its use in analytic
marginalization over waveform uncertainty, and summarize the \texttt{SEOBNRv4} model. We go on
to discuss the construction of our new uncertainty model, as well as an introduction to Bayesian \ac{PE}.
In Sec.~\ref{sec:PE_campaign} we present setup and results of
our \ac{PE} study which compares posterior distributions obtained with a number of aligned-spin
$(2,2)$ mode models, including our uncertainty model on a set of \ac{NR} surrogate signals over
parameter space. Lastly, we summarize our results and conclusions in Sec.~\ref{sec:conclusion}.

%==========================================================================
\section{Methodology}
\label{sec:methods}

In this section we discuss metrics for comparing waveforms in Sec.~\ref{sub:waveform_metrics},
summarize Gaussian process regression in Sec.~\ref{sub:gaussian_process_regression}, and how
Gaussian processes can be used to define a likelihood function which marginalizes over waveform
uncertainty in Sec.~\ref{sub:analytic_marginalization_over_waveform_uncertainty}. We discuss
the calibration of the \texttt{SEOBNRv4} waveform model against \ac{NR} simulations
in Sec.~\ref{ssec:seobnrv4_model}, and the construction of an \texttt{SEOBNRv4}-based model which 
includes waveform uncertainty in Sec.~\ref{ssec:uncertainty_model}. Finally, we introduce Bayesian
parameter estimation for compact binary coalescences in Sec.~\ref{sub:parameter_estimation}.

%==========================================================================
\subsection{Waveform metrics and parameters} % (fold)
\label{sub:waveform_metrics}

% Waveform parameters
A \ac{GW} signal emitted by an aligned-spin quasi-circular \ac{BBH} observed in the detector frame depends on 
the component masses $m_1, m_2$, dimensionless spins $\chi_i = \hat L \cdot S_i/m_i^2$ for \acp{BH} $i=1,2$ where $\hat L$ is the angular orbital momentum unit vector,
the time $t_c$ and phase $\phi_c$ of coalescence,
the inclination angle $\iota$ of the source as seen by the observer,
the luminosity distance $d_L$ and sky location $(\mathrm{ra}, \mathrm{dec})$ of the source
and the polarization angle $\psi$.
Instead of the component masses, we use the chirp mass $\mathcal{M} = M \eta^{3/5}$
and (asymmetric) mass-ratio $q = m_1/m_2 \geq 1$, where $M = m_1 + m_2$ is the total mass, 
and $\eta = q / (1 + q)^2$ is the symmetric mass-ratio.
We use the effective aligned spin parameter $\chi_\mathrm{eff} = (m_1 \chi_1 + m_2 \chi_2) / M$.

% Metrics
We decompose a gravitational waveform $h(t)$ into modes in a basis of spherical harmonics 
of spin weight -2,
\begin{equation}
  \label{eq:GWmodes}
  h_+(t) - i h_\times(t) = \sum_{\ell,m} {}_{-2}Y_{lm}(t) h_{\ell m}(\theta, \phi).
\end{equation}

The \emph{overlap} between two waveforms $h_1$ and $h_2$ can be defined as
\begin{equation}
  \label{eq:overlap}
  \mathcal{O}(h_1, h_2) := \frac{\langle h_1 | h_2 \rangle}
                                {\sqrt{\langle h_1 | h_1 \rangle \langle h_2 | h_2 \rangle}},
\end{equation}
where the noise-weighted inner product is given by
\begin{equation}
  \label{eq:inner-product}
  \langle h_1 | h_2 \rangle = 4 \mathrm{Re} \int_{f_\mathrm{low}}^{f_\mathrm{high}} 
                              \frac{\tilde h_1(f) \tilde h_2^*(f)}{S_n(f)} df
\end{equation}
where $\, \tilde {}\,$ denotes the Fourier transform, $S_n$ is the one-sided power spectral density 
\ac{PSD} of the noise, and ${}^*$ denotes complex conjugation.

We define the \emph{match} as the overlap maximized over time and phase-shifts
between the two waveforms
\begin{equation}
  \mathfrak{M}(h_1, h_2) := \max_{t_c, \phi_c} \mathcal{O}(h_1, h_2),
\end{equation}
and the \emph{mismatch} as $\overline{\mathfrak{M}}(h_1, h_2) = 1 - \mathfrak{M}(h_1, h_2)$.

The (optimal) signal-to-noise ratio (SNR) in a single \ac{GW} detector is defined as
\begin{equation}
  \rho = \sqrt{\langle h | h \rangle}.
\end{equation}
For a multi-detector network the SNR adds in quadrature, defining the network SNR, a useful
measure of how loud a \ac{GW} signal is.

% subsection waveform_metrics (end)

%==========================================================================
\subsection{Gaussian process regression} % (fold)
\label{sub:gaussian_process_regression}

\ac{GPR}\cite{Rasmussen_Williams_GPR_book, MacKay_information_theory_book} is a (nonparametric) probabilistic interpolation algorithm.
It starts from a set of training data which we assume to be pairs of intrinsic parameters
$\vec\lambda_i$ and a waveform difference $\delta h$
\begin{equation}
  \mathcal{D} = \left\{ \left(\vec\lambda_i, \; \delta h(\vec\lambda_i) \right) \right\}_{i=1}^n.
\end{equation}
(Later we will use amplitude and phase differences instead of waveform differences.)

We assume that all waveform differences in $\mathcal{D}$ are drawn from a multi-variate
Gaussian distribution and are modeled by a \ac{GP} with zero mean and \emph{covariance function}
$k(\cdot, \cdot)$. Then, it can be shown that the joint distribution of all waveform
differences in the training set and the waveform difference at a new point
$\vec\lambda \notin \mathcal{D}$ is still multi-variate Gaussian
\begin{equation}
  P\left(
    \begin{bmatrix}
    \delta h (\vec\lambda_1)\\
    \dots\\
    \delta h (\vec\lambda_n)\\
    \delta h (\vec\lambda)
    \end{bmatrix}
  \right)
  \sim
  \mathcal{N}(\mathbf{0}, \mathbf{M})
\end{equation}
with covariance matrix
\begin{equation}
  \mathbf{M} = 
  \begin{pmatrix}
  \mathbf{K}     & \mathbf{K}_*\\
  \mathbf{K}_*^T     & K_{**}
  \end{pmatrix},
\end{equation}
where the sub-matrices are given in terms of the covariance function
\begin{equation}
  (\mathbf{K})_{ij} = k(\vec\lambda_i, \vec\lambda_j), \;
  (\mathbf{K}_*)_i = k(\vec\lambda_i, \vec\lambda), \;
  K_{**} = k(\vec\lambda, \vec\lambda).
\end{equation}

For a stationary and smooth \ac{GP} the covariance function can be chosen as
\begin{equation}
  \label{eq:gpr-covariance-function}
  k(\vec\lambda_i, \vec\lambda_j) = 
  \sigma_f^2 \, \exp \left[ -\frac{1}{2} g_{ab}\,(\vec\lambda_i - \vec\lambda_j)^a \, (\vec\lambda_i - \vec\lambda_j)^b \right] + \sigma_n^2 \delta_{ij},
\end{equation}
with $g_{ab} = \mathrm{const}$, where $\sigma_f$ and $\sigma_n$ are scale factors.
We take $g_{ab}$ to be diagonal.

To train a \ac{GP} we can maximize the hyper-likelihood (or evidence for $\mathcal{D}$) 
\begin{equation}
  \log Z(\theta | \mathcal{D}) = -\frac{1}{2} (\mathbf{K}^{-1})_{ij} 
  \, \langle \delta h(\vec\lambda_i) | \delta h(\vec\lambda_j) \rangle
  - \frac{1}{2} \log \lvert \mathbf{K} \rvert
\end{equation}
over hyper-parameters $\theta = (\sigma_f, \, g_{ab}, \sigma_n)$ to find
\begin{equation}
  \hat\theta = \underset{\vec\theta}{\mathrm{argmax}}\, Z(\vec\theta | \mathcal{D}).
\end{equation}

Having determined $k(\cdot, \cdot)$ at $\hat\theta = (\hat\sigma_f, \hat g_{ab})$ one can show that
\begin{equation}
  P(\delta h(\vec\lambda) | \mathcal{D}) \propto \exp\left[ -\frac{\lVert \delta h(\vec\lambda) - \mu(\vec\lambda) \rVert^2}{2 \sigma^2(\vec\lambda)} \right]
\end{equation}
where
\begin{align}
  \mu(\vec\lambda) &=
  (\mathbf{K}_*)_i \, (\mathbf{K}^{-1})_{ij} \, \delta h(\vec\lambda_j)\\
  \sigma^2(\vec\lambda) &= K_{**} -
  (\mathbf{K}_*)_i \, (\mathbf{K}^{-1})_{ij} \, (\mathbf{K}_*)_j.
\end{align}

% subsection gaussian_process_regression (end)

\subsection{Analytic marginalization over waveform uncertainty} % (fold)
\label{sub:analytic_marginalization_over_waveform_uncertainty}

We briefly summarize the ideas of Moore \& Gair~\cite{Moore:2014pda,Moore:2015sza}.
The basic idea is to construct a prior probability distribution on the
waveform difference $\delta h(\vec\lambda) = H(\vec\lambda) - h(\vec\lambda)$,
where $H(\vec\lambda)$ is taken to be an approximate and $h(\vec\lambda)$ an
accurate waveform model (e.g. \ac{NR}).

The standard likelihood function assuming stationary Gaussian noise for the accurate model is
\begin{equation}
  L'(s|\vec\lambda) \propto \exp \left[ -\frac{1}{2} \lVert s - h(\vec\lambda) \rVert^2 \right],
\end{equation}
but in practice we only have access to the approximate model over parameter space with likelihood
\begin{equation}
  \label{eq:approx-likelihood}
  L(s|\vec\lambda) \propto \exp \left[ -\frac{1}{2} \lVert s - H(\vec\lambda) \rVert^2 \right] \approx L'.
\end{equation}

One can then define a likelihood which accounts for uncertainty in $H(\vec\lambda)$
\begin{multline}
  \mathcal{L}(s|\vec\lambda) \propto \int \mathrm{d}\left(\delta h(\vec\lambda)\right) \, P(\delta h(\vec\lambda)) \\
  \exp\left[ -\frac{1}{2} \big\lVert s - H(\vec\lambda) + \delta h(\vec\lambda) \big\rVert^2 \right].
\end{multline}
We will assume that $\delta h(\vec\lambda)$ is the realization of a \ac{GP}.
A GP furnishes a conditional probability distribution given training data $\mathcal{D}$
\begin{equation}
  \delta h(\vec\lambda) \sim \mathcal{GP}(m(\vec\lambda)=0,\, k(\vec\lambda, \vec\lambda'))
\end{equation}
or
\begin{equation}
  P(\delta h(\vec\lambda) | \mathcal{D}) \propto \exp\left[ -\frac{\lVert \delta h(\vec\lambda) - \mu(\vec\lambda) \rVert^2}{2 \sigma^2(\vec\lambda)} \right],
\end{equation}
where
\begin{equation}
  \mathcal{D} = \left\{ \left(\vec\lambda_i, \; \delta h(\vec\lambda_i) \right) \right\}_{i=1}^n.
\end{equation}

Therefore, the likelihood marginalized over waveform uncertainty is a convolution of two Gaussians
\begin{equation}
  \label{eq:marg_likelihood}
  \mathcal{L}(s|\vec\lambda) \propto \frac{1}{1 + \sigma^2(\vec\lambda)}\, \exp\left[ -\frac{1}{2} \frac{\lVert s - H(\vec\lambda) + \mu(\vec\lambda) \rVert^2}{1 + \sigma^2(\vec\lambda)} \right].
\end{equation}
Comparing Eq.\eqref{eq:marg_likelihood} with the likelihood for the approximate model Eq.~\eqref{eq:approx-likelihood} we see that the mean has been shifted by $\mu(\vec\lambda)$
and the variance has increased by $\sigma^2(\vec\lambda)$.

% subsection analytic_marginalization_over_waveform_uncertainty (end)

%==========================================================================
\subsection{The SEOBNRv4 waveform model}
\label{ssec:seobnrv4_model}

\texttt{SEOBNRv4}~\cite{Bohe:2016gbl} models quasi-circular non-precessing aligned-spin \acp{BBH}
in the effective-one-body~\ac{EOB} formalism. The binary dynamics for the inspiral and plunge is
determined by an \ac{EOB} Hamiltonian and a radiation reaction force. The final \ac{GW} waveform
combines inspiral-plunge and merger-ringdown parts of the $(2, 2)$ mode. It depends on four
calibration parameters which are then tuned to data from a set of \ac{NR} simulations.

The uncalibrated \texttt{SEOBNRv4} model $h_\mathrm{EOB}(\lambda, \theta)$ depends on two sets of parameters:
(i) three intrinsic physical parameters $\lambda = (q, \chi_1, \chi_2)$, and
(ii) four \emph{calibration parameters} $\theta = (K, d_{SO}, d_{SS}, \Delta t_\mathrm{peak}^{22})$.
The calibration parameters enter into the inspiral-plunge part of the \ac{EOB} waveform. $K$ determines the
position of the \ac{EOB} horizon and the shape of the radial potential in the strong-field region,
$d_{SO}$ and $d_{SS}$ are spin-orbit and spin-spin parameters which enter the \ac{EOB} spin mapping, and
$\Delta t_\mathrm{peak}^{22}$ is the difference between the peak of orbital frequency and the peak of radiation
in time.
Of course the model also depends on additional physical parameters listed in Sec.~\ref{sub:waveform_metrics}
in a straightforward manner; here we single out $\lambda$ because this is the subset of physical
parameters over which \ac{BBH} \ac{NR} simulations vary.

Calibration against \ac{NR} simulations fixes a mapping $\theta(\lambda)$ based on the
match between \ac{EOB} waveforms and \ac{NR} at 141 specific points $\{\lambda_i\}$ in parameter space
(see App. C of Ref.~\cite{Bohe:2016gbl} for the list of included \ac{NR} simulations, and Fig. 1 of Ref.~\cite{Bohe:2016gbl} for a graphical representation of their distribution).
The point of calibration is to increase the accuracy of the \ac{EOB} model when compared to independent \ac{NR} simulations. An accuracy measure can be defined as the mismatch
\begin{equation}
  \overline{\mathfrak{M}}(\theta) = \overline{\mathfrak{M}}(h_\mathrm{EOB}(\lambda; \theta), h_\mathrm{NR}(\lambda))
\end{equation}
but the authors of~\cite{Bohe:2016gbl} also found it beneficial to include a term
$\delta t_\mathrm{peak}^{22}(\theta)$ for the merger time difference between \ac{EOB} and \ac{NR}, after phase alignment. Given its definition, $\Delta t_\mathrm{peak}^{22}$ affects the \ac{EOB} merger time, and therefore
$\delta t_\mathrm{peak}^{22}(\theta)$.
The mismatch and merger time difference are combined in the definition of a Gaussian likelihood function
\begin{multline}
  \label{eq:cal-likelihood}
  p(\lambda_i | \theta) \propto\\
  \exp \left[ - \frac{1}{2} \left(\frac{{\overline{\mathfrak{M}}}_\mathrm{max}(\theta)}{\sigma_\mathfrak{M}} \right)^2 
  - \frac{1}{2} \left( \frac{\delta t^{22}_\mathrm{peak}(\theta)}{\sigma_t} \right)^2
  \right],
\end{multline}
where $\overline{\mathfrak{M}}_\mathrm{max}(\theta)$ is the maximum mismatch over total masses
between $10$ and $200 M_\odot$. The standard deviations are chosen as $\sigma_{\mathfrak{M}} = 0.01$
and $\sigma_t = 5 M$.

\begin{figure}[!htbp]
  \centering
    \includegraphics[width=.45\textwidth]{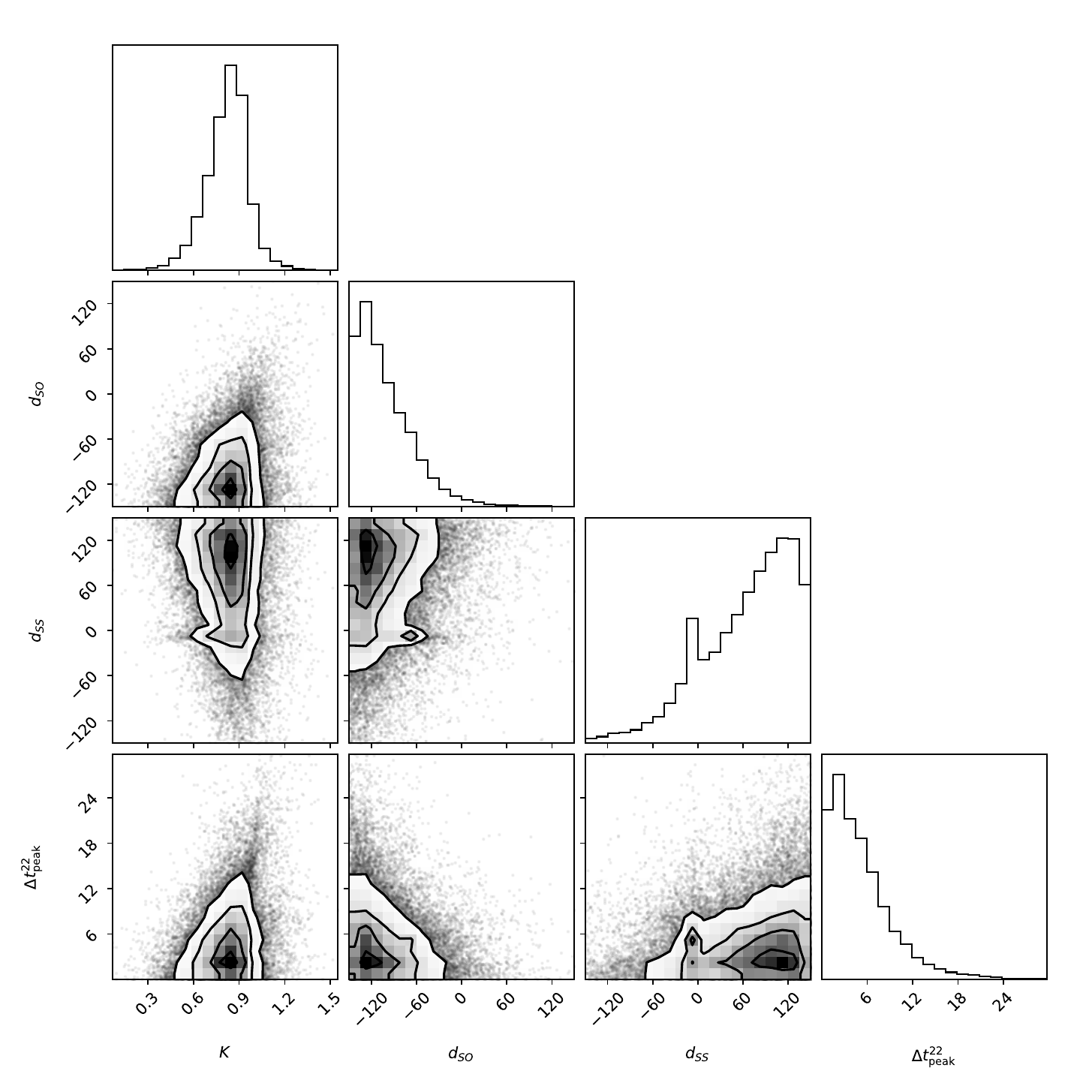}
  \caption{
    A corner plot of an example calibration posterior distribution $p(\theta | \lambda_i)$
    for and SXS simulation at $q=2, \chi_1 = 0, \chi_2 = -0.6$.
  }
  \label{fig:figures_PRIVATE_BBH_0089_calibration_posterior}
\end{figure}

The calibration procedure involves the following steps:
(i) For each $\lambda_i$ determine the \emph{calibration posterior} distribution
$p(\theta | \lambda_i)$ assuming a flat prior in $\theta$
using an MCMC sampler. See Fig.~\ref{fig:figures_PRIVATE_BBH_0089_calibration_posterior} for an example.
(ii) Discard calibration posterior samples with $\overline{\mathfrak{M}}_\mathrm{max} > 1 \%$ and $|\delta t^{22}_\mathrm{peak}| > 5 M$.
(iii) Remove secondary modes from the posteriors $p(\theta | \lambda_i)$

Finally, a polynomial fit is performed to the means of the calibration posteriors $\overline{\theta}(\lambda_i)$ to obtain a mapping $\theta(\lambda)$ throughout physical parameter space as discussed in Ref.~\cite{Bohe:2016gbl}.
With this prescription it was found that the maximum mismatch of the resulting calibrated \texttt{SEOBNRv4} 
model against the available \ac{NR} waveforms was less than $1\%$. Detailed mismatch comparisons against NR are shown in~\cite{Bohe:2016gbl}.

%==========================================================================
\subsection{Model for waveform uncertainty in SEOBNRv4}
\label{ssec:uncertainty_model}

We introduce a novel method to build an effective-one-body model which includes waveform
uncertainty based on posterior distributions of EOB calibration parameters given a set
of NR simulations. This allows us to replace a usual deterministic waveform model $h(\lambda)$
with a probabilistic waveform model $p(h | \lambda)$. Our method uses a computationally
efficient ansatz in terms of amplitude and phase deviations. These deviations are computed 
between the calibrated SEOBNRv4 model and samples from an uncalibrated SEOBNRv4 conditioned
on calibration posteriors. We first determine the distribution of deviations at each NR 
calibration point $\lambda$, and subsequently interpolate the distributions over the binary
parameter space.

\subsubsection{Construction of calibration error model} % (fold)
\label{ssub:construction_of_calibration_error_model}

We make the following ansatz for the calibration error (CE) model to achieve a compressed
frequency domain representation of the waveform uncertainty arising in calibration of 
SEOBNRv4 (here abbreviated as ``EOB'') against NR simulations:
\begin{equation}
  \label{eq:CE-ansatz}
  \tilde h_\mathrm{CE}(\lambda; f) = \left[1 + \delta A(\lambda; f)\right] \, e^{i \delta\phi(\lambda; f)}
  \times \tilde h_\mathrm{EOB}(\lambda; f),
\end{equation}
where $\delta A$ and $\delta \phi$ are relative amplitude and absolute phase deviations.
Expanding $\tilde h_\mathrm{EOB} = A_\mathrm{EOB} \, e^{i \phi_\mathrm{EOB}}$ the deviations
can be computed as
\begin{align}
  \delta A(f)   &= A_\mathrm{CE}(f) / A_\mathrm{EOB}(f) - 1\\
  \delta\phi(f) &= \phi_\mathrm{CE}(f) - \phi_\mathrm{EOB}(f).
\end{align}

At each NR calibration point $\lambda_i$ we compute the distribution of deviations
\begin{equation}
  \label{eq:actual_differences}
  p\left((\delta A, \delta\phi) | \lambda_i\right)
\end{equation}
given the uncalibrated SEOBNRv4 model $h_\mathrm{EOB}(\lambda, \theta)$ and calibration
posterior samples $\theta_k \sim p(\theta | h_\mathrm{NR}(\lambda_i|))$. I.e. we set
$A_\mathrm{CE}^{i,k} = A_\mathrm{EOB}(\lambda_i, \theta_k)$,
$\phi_\mathrm{CE}^{i,k} = \phi_\mathrm{EOB}(\lambda_i, \theta_k)$.
At this point we only retain the deviations at a log-spaced set of $J=10$ frequency nodes,
defining 
\begin{align}
  \delta   A_j &= \delta A(f_j)\\
  \delta\phi_j &= \delta\phi(f_j),
\end{align}
where the nodes $\{ f_j \}_{j=1}^J$ are spaced uniformly in logarithmic frequency in the interval 
$\approx [20, 406]$ Hz, where the upper cutoff is determined from a maximum geometric frequency $Mf_\mathrm{max} = 0.1$ at a total mass of $50 M_\odot$.
We chose to use log-frequency spacing based on the note~\cite{SplineCalMarg-T1400682} on calibration uncertainty. This choice, along with the choice of ten nodes, seemed to capture the waveform uncertainty morphology reasonably well, particularly the increased structure at lower frequencies.
For ease of notation, we drop the discrete frequency index in the following discussion.

Collecting the discrete deviations in a vector, we approximate these distributions with
a multivariate normal ansatz
\begin{equation}
  \label{eq:mvn_ansatz}
  p\left((\vec\delta A, \vec\delta\phi) | \lambda_i|\right) \approx \mathcal{N}_{2J}(\vec\mu(\lambda_i), \Sigma(\lambda_i)),
\end{equation}
with mean vectors $\vec\mu(\lambda_i) \in \mathbb{R}^{2J}$ and covariance matrices $\Sigma(\lambda_i) \in \mathbb{R}^{2J \times 2J}$.

Finally, we use \ac{GPR}~\cite{Rasmussen_Williams_GPR_book} to interpolate
this information over the parameter space, generalizing to arbitrary $\lambda$ near
the calibration points
\begin{equation}
  \left(\vec\mu(\lambda_i), \Sigma(\lambda_i)\right) \mapsto \left(\vec\mu(\lambda), \Sigma(\lambda)\right).
\end{equation}
To do this, it is advantageous to perform a Cholesky decomposition of each covariance matrix
in order to preserve its positive semi-definiteness
\begin{equation}
  \Sigma(\lambda_i) = L(\lambda_i)^T L(\lambda_i),
\end{equation}
and we perform GPR interpolation on the set of $\{ \vec\mu(\lambda_i), L(\lambda_i) \}$,
where the lower triangular matrices $L$ are suitably arranged into a vector.
We only use the GPR means in the uncertainty model. In this implementation the stochasticity
of the model comes entirely from the modeling of the covariances $\Sigma(\lambda)$.

% subsubsection construction_of_calibration_error_model (end)

\subsubsection{Evaluation of calibration error model} % (fold)
\label{ssub:evaluation_of_calibration_error_model}

The CE model allows for rapid draws of amplitude and phase deviation samples
at a fixed parameter space point $\lambda$
\begin{equation}
  \left( \{ \vec\delta A \}, \{ \vec\delta\phi \}  \right) \sim p \left( (\delta A, \delta\phi) | \lambda| \right).
\end{equation}
These samples can be plugged into the CE model ansatz Eq.~\eqref{eq:CE-ansatz} to 
obtain a distribution of CE waveforms $p(h_\mathrm{CE} | \lambda|)$.

The GPR evaluation is achieved by (i) drawing multi-variate normal vectors
\begin{equation}
  \label{eq:epsilon_normal}
  \vec\epsilon_A, \vec\epsilon_\phi \sim \mathcal{N}_J(0, I),
\end{equation}
(ii) evaluating the GPR fits for amplitude and phase mean vectors $\vec\mu$
and the decomposed covariance matrices $L$ at point $\lambda$,
(iii) computing the amplitude and phase deviations
\begin{align}
  \label{eq:modeled_differences}
  \vec\delta A   &= \vec\mu_A(\lambda) + L_A(\lambda) \vec\epsilon_A\\
  \vec\delta\phi &= \vec\mu_\phi(\lambda) + L_\phi(\lambda) \vec\epsilon_\phi,
\end{align}
(iv) constructing cubic splines for $\delta A(f)$ and $\delta\phi(f)$, and
evaluating the CE waveform Eq.~\eqref{eq:CE-ansatz} on the desired frequency grid.
We refer to the uncertainty model as \texttt{SEOBNRv4CE} and, the mean uncertainty model
with $\vec\epsilon = 0$, as \texttt{SEOBNRv4CE0}.

\begin{figure}[!htbp]
  \centering
    \includegraphics[width=.45\textwidth]{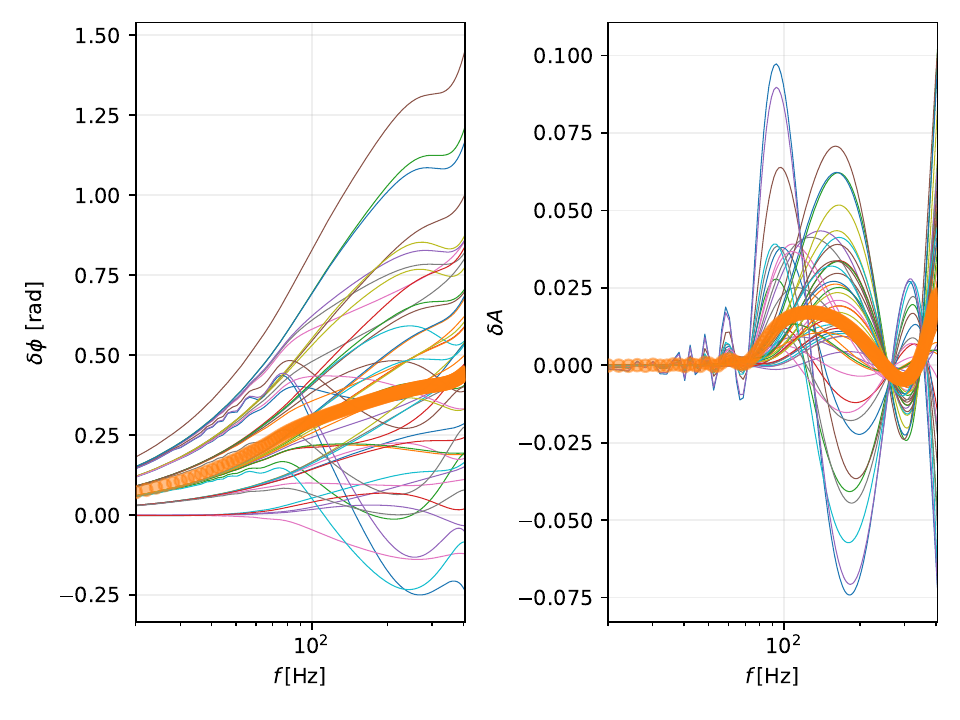}
    \includegraphics[width=.45\textwidth]{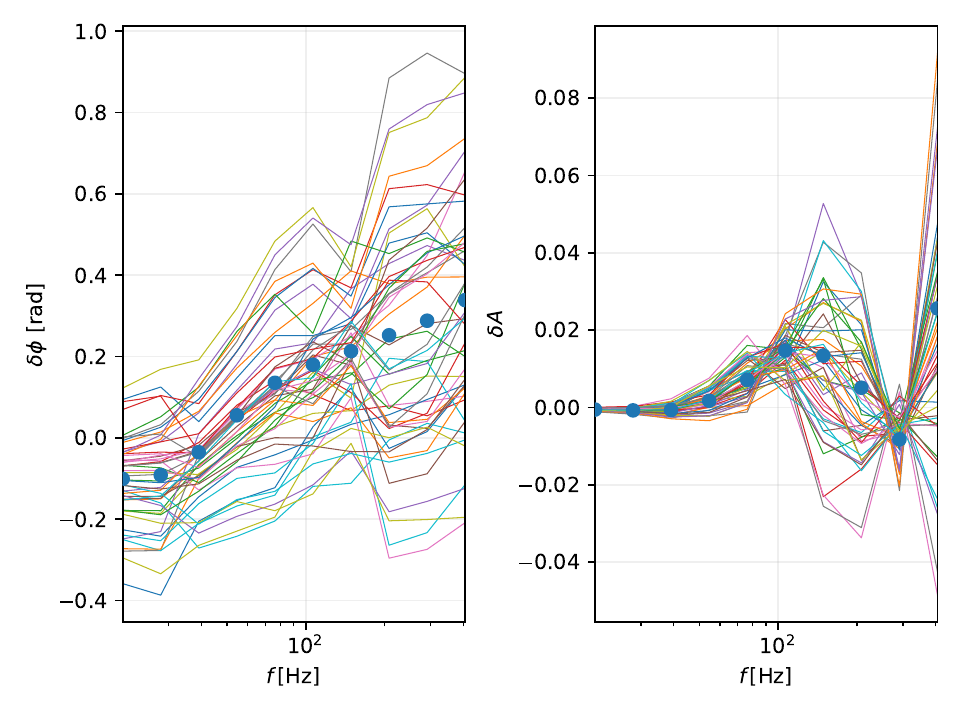}
  \caption{
    Actual (top) and predicted (bottom) amplitude and phase differences for 
    \texttt{SXS\_BBH\_0166} $q = 6.0$, $\chi_1 = 0.0$, $\chi_2 = 0.0$.
    Actual differences are samples from the distribution defined in Eq.\eqref{eq:actual_differences}
    while the differences predicted by the uncertainty model are given by Eq.\eqref{eq:modeled_differences}.
    The mean differences are indicated by filled orange and blue circles in the top and bottom panels, respectively.
  }
  \label{fig:figures_actual_and_modeled_differences}
\end{figure}

\begin{figure}[!htbp]
  \centering
    \includegraphics[width=.45\textwidth]{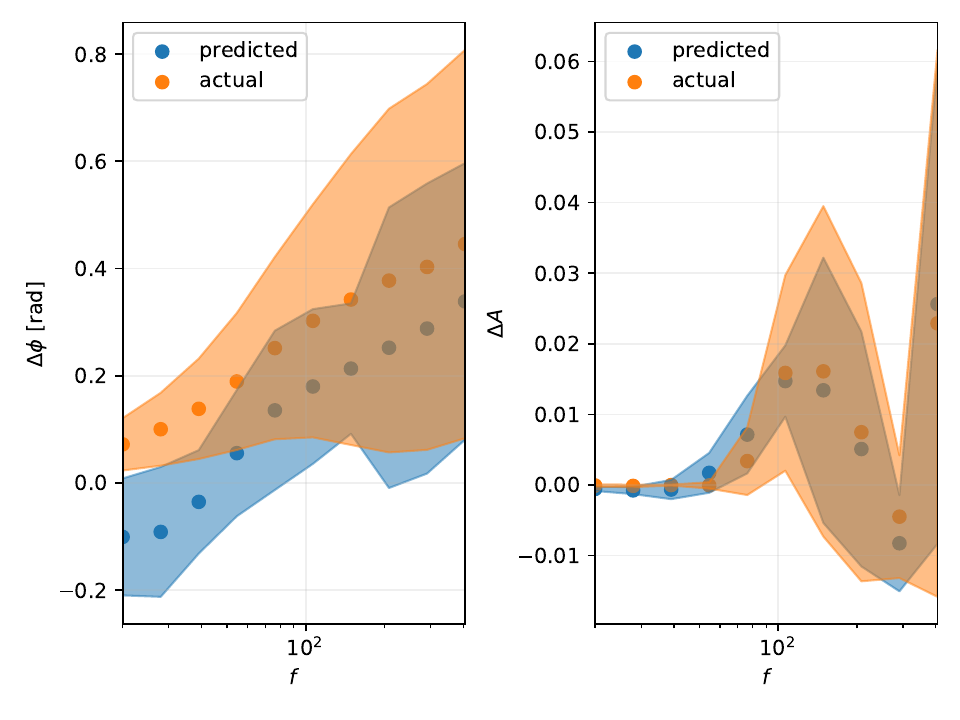}
    \includegraphics[width=.45\textwidth]{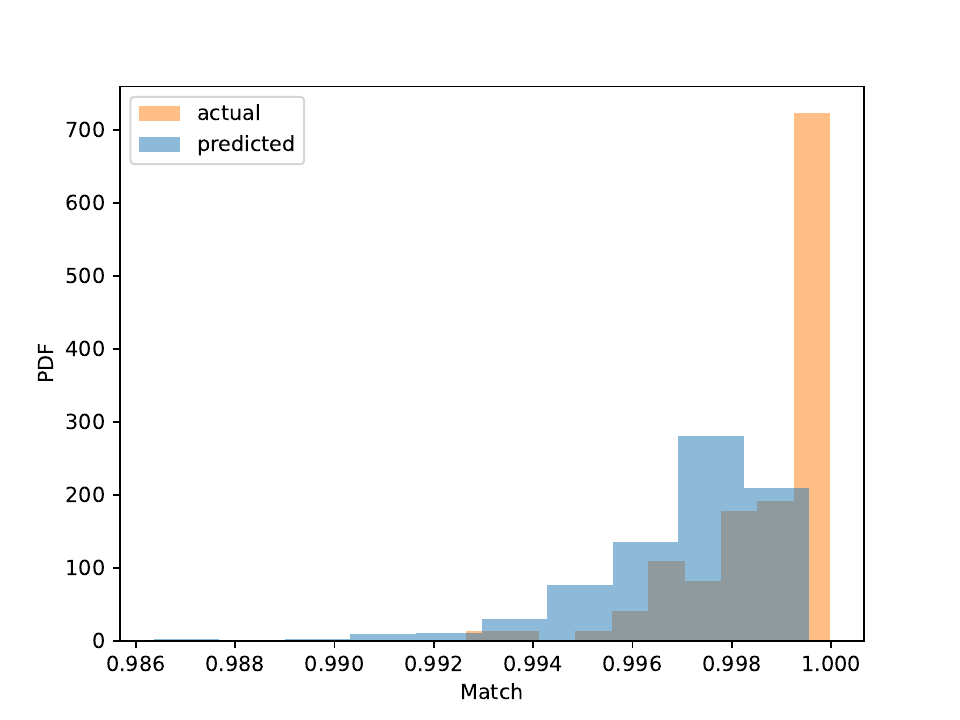}
  \caption{
    Top: comparison of actual and predicted means and $1-\sigma$ uncertainty (neglecting correlations
    between nodes at different frequencies) for the same configuration shown in
    Fig.~\ref{fig:figures_actual_and_modeled_differences}.
    Bottom: matches between uncalibrated SEOBNRv4 at draws from calibration posterior and SEOBNRv4 (actual)
    and between the uncertainty model and SEOBNRv4 (predicted).
  }
  \label{fig:figures_SXS_BBH_0166_q6_0_0_means_comparison}
\end{figure}

Fig.~\ref{fig:figures_actual_and_modeled_differences} shows the approximation of
amplitude and phase differences by the uncertainty model for a simulation in the test set.
They generally follow the same shape in frequency. As shown in
Fig.~\ref{fig:figures_SXS_BBH_0166_q6_0_0_means_comparison}, while the means are somewhat
offset, the bulk of the uncertainty regions overlap. A linear difference in the phase may also
be compensated by time and phase shifts. Indeed, the second panel of this figure shows that
while the predicted match distribution extends to lower match values, it overlaps well with
the actual match distribution.

\begin{figure}[!htbp]
  \centering
    \includegraphics[width=.45\textwidth]{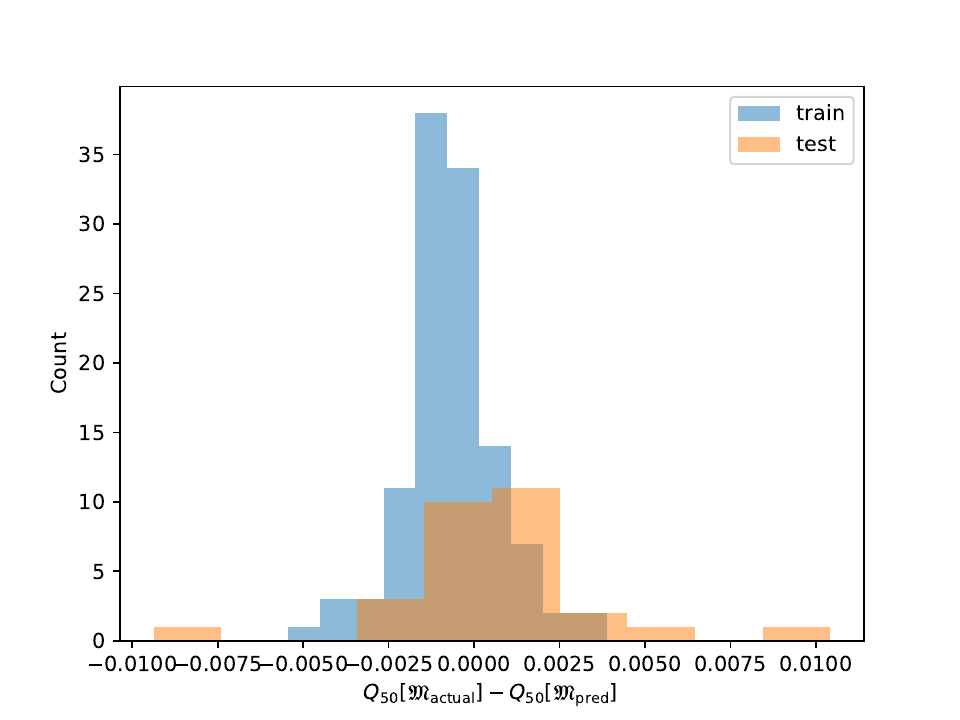}
  \caption{
    Difference between medians of matches between uncalibrated SEOBNRv4 at draws from calibration posterior and SEOBNRv4 ($\mathfrak{M}_\mathrm{actual}$) and matches between the uncertainty model and SEOBNRv4 ($\mathfrak{M}_\mathrm{pred}$) for both the training and test \ac{NR} configurations. 
  }
  \label{fig:figures_median_match_difference}
\end{figure}

Fig.~\ref{fig:figures_median_match_difference} indicates that both for training and test configurations in some cases the uncertainty model does better than the calibration posteriors, and in other cases it is the opposite. The spread of the difference is somewhat larger for the test cases which is to be expected because the uncertainty model only approximates and interpolates the calibration posterior information. A similar picture is seen when comparing the full match distributions.

% subsubsection evaluation_of_calibration_error_model (end)

\subsubsection{Interpretation of calibration error model} % (fold)
\label{ssub:interpretation_of_calibration_error_model}

Parametrized by $\vec\epsilon_A, \vec\epsilon_\phi$, the \texttt{SEOBNRv4CE} model returns
a distribution of waveforms with amplitude and phase deviations around the
means $\vec\mu_A(\lambda), \vec\mu_\phi(\lambda)$ which arise from interpolation
of the mean deviations in amplitude and phase between the calibrated \texttt{SEOBNRv4} and
the uncalibrated model conditioned on NR-calibration information.

For computational efficiency we use the frequency domain \ac{ROM} of
\texttt{SEOBNRv4}~\cite{Purrer:2014fza,Purrer:2015tud,Bohe:2016gbl} which closely approximates the time domain \texttt{SEOBNRv4} model:

\begin{multline}
  \label{eq:seobnrv4ce_model}
  \tilde h_\mathrm{CE}(\lambda, \vec\epsilon; f) = \tilde h_\mathrm{ROM}(\lambda; f)
  \left[ 1 + \delta A(\lambda, \vec\epsilon; f) \right] \\
  \exp(i \delta\phi(\lambda, \vec\epsilon; f)).
\end{multline}

\begin{figure}[!htbp]
  \centering
    \includegraphics[width=.45\textwidth]{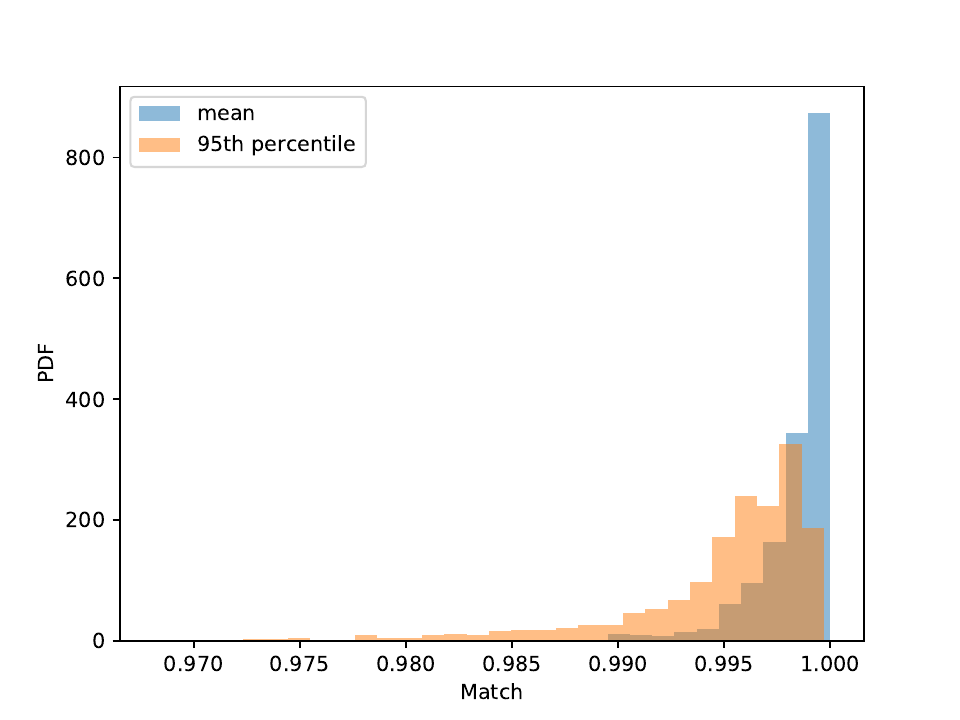}
  \caption{
    Histogram of matches between \texttt{SEOBNRv4} and the \texttt{SEOBNRv4CE} uncertainty model, either assuming
    mean deviations ($\vec\epsilon = 0$, \texttt{SEOBNRv4CE0}) or the 95th percentile of the match
    distribution assuming Eq.~\eqref{eq:epsilon_normal}.
  }
  \label{fig:figures_SEOBNRv4CE_SEOBNRv4_matches}
\end{figure}

For the sake of comparison we also consider a deterministic waveform model where we
assume a neutral or mean calibration $\vec\epsilon = 0$ which we refer to as \texttt{SEOBNRv4CE0}.
Because of the normal approximation and GPR interpolation of waveform differences used in
\texttt{SEOBNRv4CE}, $\vec\epsilon = 0$ does not exactly correspond to $\delta A, \delta\phi = 0$
over the parameter space. Therefore \texttt{SEOBNRv4CE0} is close, but not exactly
the same as the calibrated \texttt{SEOBNRv4} (or its ROM approximation).
Indeed, Fig.~\ref{fig:figures_SEOBNRv4CE_SEOBNRv4_matches} shows that the mean uncertainty model
(\texttt{SEOBNRv4CE0}) is very close to \texttt{SEOBNRv4} and deviations reach about $\sim 2\%$
in mismatch at the 95th percentile of the match distribution.

% subsubsection interpretation_of_calibration_error_model (end)

%==========================================================================
\subsection{Parameter estimation} % (fold)
\label{sub:parameter_estimation}

The main goal of parameter estimation is to find the posterior distribution
$p\left(\Lambda | \{d_k\}, \mathcal{H}\right)$ of the \ac{BBH} parameters $\Lambda$ using Bayes' theorem,
\begin{equation}
  p\left(\Lambda | \{d_k\}, \mathcal{H}\right) = \frac{\mathcal{L}(\{d_k\} | 
  \Lambda, \mathcal{H}) \pi(\Lambda | \mathcal{H})}{\mathcal{Z}(\{d_k\} | \mathcal{H})},
  \label{eq:bayes_theorem}
\end{equation}
where $\mathcal{H}$ is the signal hypothesis and $d_{i}(t)$ is the time domain strain data 
in interferometer $i$ of the LVK network.

The signal and template waveforms are projected on the interferometric \ac{GW} detectors
and we compute the strains from the \ac{GW} polarizations and detector pattern functions~\cite{Chatziioannou:2024hju}
\begin{equation}
  h(t; \Lambda) = F_+(\mathrm{ra}, \mathrm{dec}, \psi) h_+(t; \Theta) + F_\times(\mathrm{ra}, \mathrm{dec}, \psi) h_\times(t; \Lambda)
\end{equation}

The prior probability distribution $\pi(\Lambda | \mathcal{H})$ in Eq.~\eqref{eq:bayes_theorem}
encapsulates our prior knowledge of the distribution of $\Lambda$ and $\mathcal{L}(\{d_k\} |
  \Lambda, \mathcal{H})$ is the likelihood of a observing the data $d$ given an assumed
  waveform model $h^{\mathcal{H}}(t; \Lambda)$.
The Whittle likelihood for stationary Gaussian noise for a network of detectors is~\cite{Thrane:2018qnx,Chatziioannou:2024hju}
\begin{equation}
  \mathcal{L}(\{d_i\}|\Lambda, \mathcal{H}) \propto \prod_{i}  \mathrm{exp} \left( -\frac{1}{2}
  \langle d_i - h^{\mathcal{H}}_i(\Lambda) | d_i - h^{\mathcal{H}}_i(\Lambda) \rangle\right),
  \label{eq:likelihood_func}
\end{equation}
where $\langle \cdot | \cdot \rangle$ is the noise-weighted inner product of Eq.~\eqref{eq:inner-product}.

The normalizing constant
\begin{equation}
  \mathcal{Z}(\{ d_i\} | \mathcal{H}) = \int d\Lambda \, \pi(\Lambda | \mathcal{H}) 
  \mathcal{L}(\{d_i\} | \Lambda, \mathcal{H})
\end{equation}
in Eq. \ref{eq:bayes_theorem} is called evidence and can be used for model selection.
The marginal posterior probability distribution on a subset of binary parameters $\vec{\lambda_{A}}$,
where $\Lambda = \{\vec{\lambda_A}, \vec{\lambda_B}\}$, is defined as,
\begin{equation}
  p(\vec{\lambda_A} | \{d_i\}, \mathcal{H}) = \int_{\vec{\lambda_B}} \, d\vec{\lambda_B} p(\vec{\lambda_A} | 
  \{d_i\}, \mathcal{H}).
\end{equation}

In this study we use the parameter estimation code \texttt{Bilby}~\cite{Ashton:2018jfp,bilby_pipe_paper}
and the nested sampling algorithm \texttt{Dynesty}~\cite{Speagle:2019dynesty,sergey_koposov_2024_12537467}.
We choose an HLV network at advanced LIGO and advanced Virgo design sensitivity~\cite{aLIGODesignNoiseCurve,adVirgoDesignNoiseCurve}.
We use an average \emph{zero noise} realization in this study to focus on waveform systematics and avoid the effect of random 
Gaussian noise realizations on posteriors. With this assumption biases observed in posterior distributions can arise from 
discrepancies in signal and template waveform models or prior effects.

We choose a uniform prior on chirp mass $\mathcal{M} \sim \mathcal{U}[25, 50] M_\odot$ and mass-ratio $1/q \sim \mathcal{U}[0.125, 1]$, 
uniform priors for the dimensionless aligned spins of the component \acp{BH} $\chi_i \sim \mathcal{U}[-0.8, 0.8]$ with 
limits arising from the calibration region of \texttt{NRHybSur3dq8}~\cite{Varma:2018mmi}. We assume an isotropic prior for 
the location of the \ac{GW} signal on the sky and a distance prior corresponding to a homogenous rate density in the nearby 
Universe.
Finally, we choose $\cos\theta_\mathrm{JN} \sim \mathcal{U}[0, \pi]$ and a uniform prior in
geocentric time $t_c \sim \mathcal{U}[-0.1, 0.1]$ around the geocentric time of the injected signal.

When we perform inference with the \texttt{SEOBNRv4CE} uncertainty model we choose unit normal priors for the components 
of $\epsilon$ following Eq.~\eqref{eq:epsilon_normal}. We typically marginalize over these uncertainty parameters.

%==========================================================================
\section{Parameter estimation with waveform uncertainty}
\label{sec:PE_campaign}

Here we test the effectiveness of the \texttt{SEOBNRv4CE} uncertainty model we introduced in Sec.~\ref{ssec:uncertainty_model} by conducting a parameter 
estimation campaign using mock \ac{GW} data. Our primary objective is to study 
systematic biases induced by the \texttt{SEOBNRv4} waveform model and evaluate how well the \texttt{SEOBNRv4CE} model mitigates these effects.

%==========================================================================
\subsection{Parameter estimation setup}
\label{sub:PE_setup}

In this study we use the waveform models
\texttt{SEOBNRv4\_ROM}~\cite{Purrer:2014fza,Purrer:2015tud,Bohe:2016gbl}, and
\texttt{NRHybSur3dq8}~\cite{Varma:2018mmi},as available in the LALSuite~\cite{lalsuite} 
software package, in addition to the new uncertainty model \texttt{SEOBNRv4CE} we developed 
in Sec.~\ref{ssec:uncertainty_model}.

\texttt{SEOBNRv4\_ROM}~\cite{Bohe:2016gbl} is a frequency domain reduced order model (ROM) 
of the time domain \texttt{SEOBNRv4} model~\cite{Bohe:2016gbl} using the methodology 
developed in~\cite{Purrer:2014fza,Purrer:2015tud}. It provides an accurate and computationally
much faster representation of \texttt{SEOBNRv4} waveforms. This model includes the $(2,\pm 2)$
modes for non-precessing binary black holes and can be used for a wide range in mass-ratio 
and BH spin magnitudes up to maximal spin.
The new uncertainty model we develop in this study is based on \texttt{SEOBNRv4\_ROM} and 
the set of calibration posteriors $p(\theta | \lambda_i)$ obtained for 
\texttt{SEOBNRv4}. While \texttt{SEOBNRv4} is not the newest SEOBNR model \cite{Pompili_Lorenzo:SEOBNRv5} 
the focus of this paper is more on showcasing an efficient method of marginalizing over waveform systematics 
in parameter estimation.

We use the \ac{NR}-hybrid surrogate model \texttt{NRHybSur3dq8}~\cite{Varma:2018mmi}
as an independent and highly accurate comparison waveform for parameter estimation studies. 
\texttt{NRHybSur3dq8} models \acp{BBH} with aligned spin magnitudes 
$|\chi_i| \leq 0.8$ and mass-ratios smaller than $q \leq 8$ and includes a set of higher 
harmonics up to $\ell \sim 4$. Here we only include the dominant $(2,\pm 2)$ mode of
\texttt{NRHybSur3dq8}, so that all models can be compared on an equal footing.

We represent all signals by \texttt{NRHybSur3dq8} to act as a stand-in for \ac{NR} simulations because of its superior accuracy. We assume a zero noise realization and an HLV network at design sensitivity.
Analyses follow the Bayesian framework laid out in Sec. \ref{sub:parameter_estimation}
to estimate the posterior distribution $p(\Lambda|d, \mathcal{H})$ on BBH parameters.
To explore how well the method works over parameter space we analyze a grid of signals defined as the Cartesian product of binary configurations: mass ratio $q = \{1.25, 4, 8\}$,
aligned spin magnitude $\chi_1=\chi_2=\{-0.5, 0, 0.5\}$, luminosity distance $d_\mathrm{L}=\{205, 410, 820, 1640\}$ Mpc. Each signal is analyzed for four template waveform models: \texttt{NRHybSur3dq8}, \texttt{SEOBNRv4}, \texttt{SEOBNRv4CE}, and \texttt{SEOBNRv4CE0}.

Remaining signal parameters are kept fixed for each configuration:
chirp mass $\mathcal{M} = 32.0499 M_\odot$, inclination angle $\iota = 2.83701491$ rad,
polarization angle $\psi = 1.42892206$~rad, 
coalescence phase $\phi_{c} = 1.3$ rad, coalescence time $t_c = 1126259462.0$ sec,
right ascension $ \mathrm{ra} = -1.26157296$~rad, and declination $\mathrm{dec} = 1.94972503$~rad.
The prior distributions are provided in Sec. \ref{sub:parameter_estimation}, and unless otherwise noted,
we marginalize over the $\epsilon$ parameters.

%==========================================================================
\subsection{Parameter estimation results}
\label{sub:PE_results}

In this section we discuss the results of the Bayesian analysis following the setup laid out in 
Sec \ref{sub:PE_setup}. 
We focus on the effect of waveform systematics on key intrinsic parameters chirp mass $\mathcal{M}$,
mass ratio $q$, and effective spin $\chieff$. 
To estimate the statistical error for a given parameter $\lambda$, we calculate the standard deviation 
$\sigma_{\lambda}$ of the one-dimensional marginal posterior distribution. To quantify how far the recovered
posterior deviates from the true (injected) value $\lambda_\mathrm{tr}$, we calculate the bias as 
$\Delta \lambda = |\lambda_\mathrm{med} - \lambda_\mathrm{tr}|$, where $\lambda_\mathrm{med}$ is the median of the 
posterior distribution. Then, the normalized bias is obtained by taking the ratio 
$\Delta \lambda /\sigma_{\lambda}$.

\subsubsection{Evolution of chirp mass posterior with SNR}

\begin{figure*}[!htbp]
  \centering
  \begin{minipage}{0.45\textwidth}
    \includegraphics[width=\linewidth]{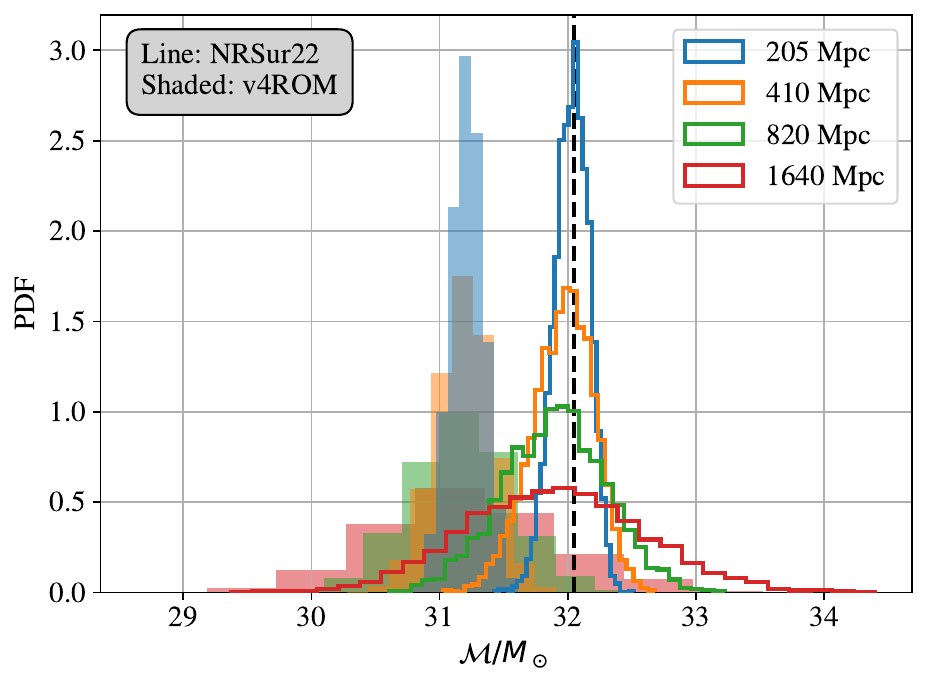}
  \end{minipage}
  \hfill
  \begin{minipage}{0.45\textwidth}
    \includegraphics[width=\linewidth]{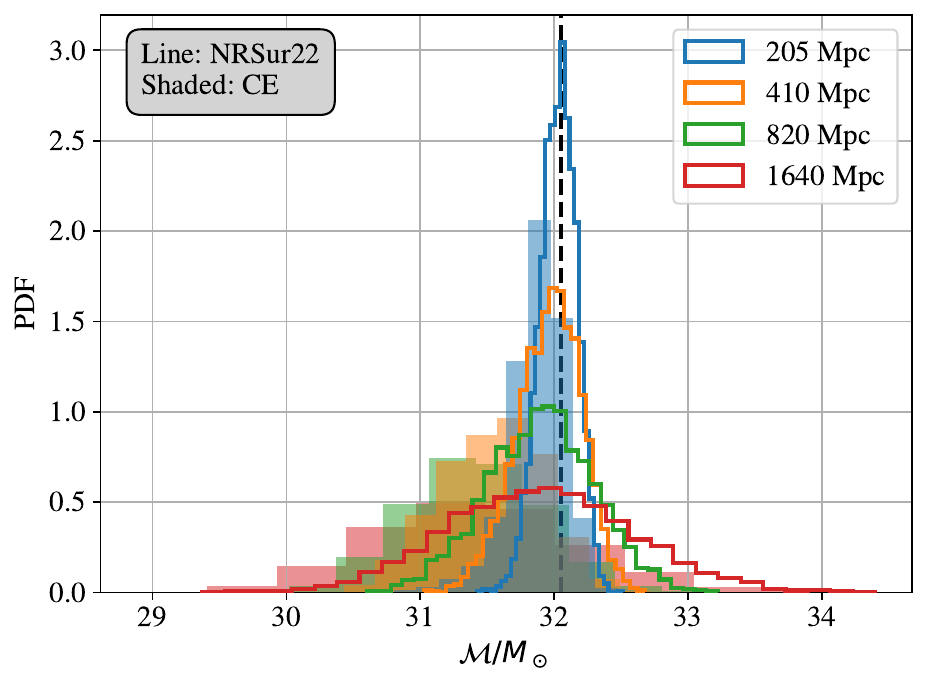}
  \end{minipage}
  \caption{The evolution of the chirp mass posterior as a function of luminosity distance is presented 
  for a binary configuration with mass ratio  $q = 4.0$ and spin magnitudes  $\chi_1 = \chi_2 = 0.5$. 
  The corresponding signal-to-noise ratios (SNRs) for luminosity distances  $d_\mathrm{L}$ = 205  Mpc, 
  410 Mpc, 820 Mpc, and 1640 Mpc are 217, 108, 54, and 27, respectively. The solid line illustrates 
  the posterior recovered using \texttt{NRHybSur3dq8}. In the left plot, the shaded area represents 
  the posterior obtained with \texttt{SEOBNRv4}, while the right plot shows the posterior distribution 
  from \texttt{SEOBNRv4CE}. We observe significant bias in the \texttt{SEOBNRv4} posterior at higher 
  SNRs, whereas much of this bias is mitigated in the \texttt{SEOBNRv4CE} model. However, the 
  \texttt{SEOBNRv4CE} posterior distribution is considerably broader.}
  \label{fig:marginal_chirp_mass}
\end{figure*}

Fig.~\ref{fig:marginal_chirp_mass} shows the evolution of the marginal chirp mass posterior
at various distances $d_\mathrm{L}$. We observe that as $d_\mathrm{L}$ decreases the distributions gets 
narrower. This behavior is expected, as the SNR is proportional to $1/d_L$. Increased SNR leads to 
a reduction in statistical error as higher SNR improves the precision in parameter estimation. 
Because the signal is chosen as \texttt{NRHybSur3dq8}, the posterior distribution obtained 
with this model peaks around the true chirp mass value. The posterior obtained using 
\texttt{SEOBNRv4} is significantly biased at higher SNR. Meanwhile, the posteriors obtained with 
\texttt{SEOBNRv4CE} show less bias compared to \texttt{SEOBNRv4} at the expense of being 
less precise. At lower SNR, the posteriors from \texttt{SEOBNRv4CE} have similar widths to 
those from \texttt{NRHybSur3dq8}, but at higher SNR, the \texttt{SEOBNRv4CE} posteriors are 
considerably wider.

%--------------------------------------------------------------------------
%          Chirp Mass Posterior Distribution Evolution with SNR
%--------------------------------------------------------------------------

\begin{figure}[!hbtp]
  \includegraphics[width=0.45\textwidth]{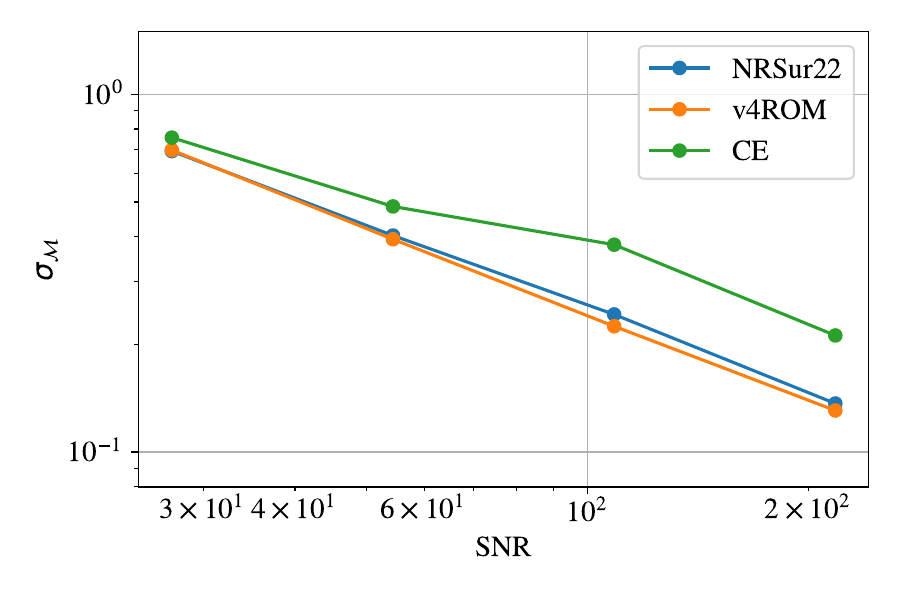}
  \caption{
    Standard deviation of chirp mass posterior as a function of SNR for the configuration
    $q = 4$, $\chi_1 = 0.5$, $\chi_2 = 0.5$ also shown in Fig.~\ref{fig:marginal_chirp_mass}
    for waveform models \texttt{NRHybSur3dq8}, \texttt{SEOBNRv4}, and \texttt{SEOBNRv4CE}.}
  \label{fig:snr_vs_sigma}
\end{figure}

In Fig.~\ref{fig:snr_vs_sigma}, we further analyze statistical uncertainties represented by the
standard deviation $\sigma$ of the one dimensional marginal posterior distribution as a function
of SNR. We observe that the standard deviation of the \texttt{SEOBNRv4CE} chirp mass posterior
is larger than that of the \texttt{NRHybSur3dq8} posterior. Moreover, the disparity between the
chirp mass standard deviations increases as the SNR rises. We also observe that the
\texttt{SEOBNRv4} posterior has similar standard deviation as the \texttt{NRHybSur3dq8} posterior.
Assuming that the joint posterior can be approximated by a multivariate Gaussian distribution at
high SNR, we expect the posterior distributions for models which do not include waveform uncertainty
to follow an inverse power law with slope $-1$. By fitting the data to the relation 
$\sigma \propto \rho^{-\alpha}$, we obtain $\alpha = -0.77$ for \texttt{NRHybSur3dq8},
$\alpha = -0.80$ for \texttt{SEOBNRv4}, and $\alpha = -0.58$ for \texttt{SEOBNRv4CE0}.
While we do not recover the ideal scaling for \texttt{NRHybSur3dq8} and \texttt{SEOBNRv4},
we do see that the exponent is less negative for \texttt{SEOBNRv4CE0} that the other models,
reflecting the broader \texttt{SEOBNRv4CE0} posteriors -- see Sec.~\ref{ssub:idealized_scaling}
for an analysis in a more idealized setting.
This observation is consistent with posterior broadening in the Moore \& Gair
formalism\cite{Moore:2014pda}, where the marginalized likelihood distribution over waveform
uncertainty in Eq.~\ref{eq:marg_likelihood} is wider compared to the standard likelihood
function in Eq.~\ref{eq:approx-likelihood}. Because \texttt{SEOBNRv4CE0} is cast as a model
of amplitude and phase deviations rather than overall waveform differences we cannot directly
access the increase in variance $\sigma^2(\lambda)$ arising from marginalization over waveform
uncertainty.

%--------------------------------------------------------------------------
%                  SNR vs Sigma - q=1.25, chi1=chi2=0.0
%--------------------------------------------------------------------------

\subsubsection{Parameter correlations}

%--------------------------------------------------------------------------
%              Corner Plot - q=4, \chieff = 0.5, \chieff=0
%--------------------------------------------------------------------------
\begin{figure*}[!hbtp]
  \centering
  \subfloat[$q = 4.0$, $\chi_1 = \chi_2 = 0.0$, $d_\mathrm{L}$ 205 Mpc]{
    \includegraphics[width=0.48\textwidth]{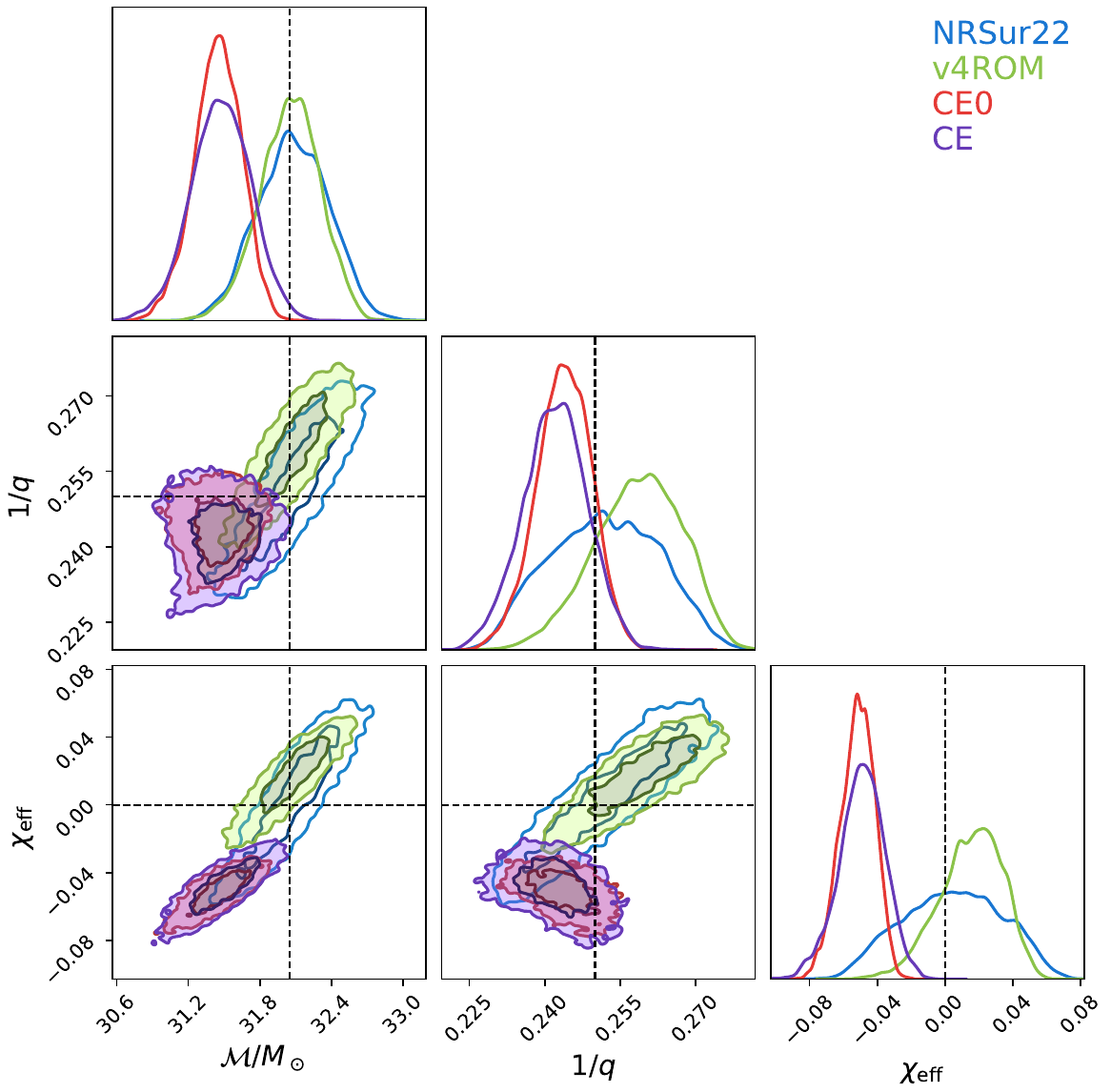}
  }
  \hfill
  \subfloat[$q=4.0$, $\chi_1 = \chi_2 = 0.5$, $d_\mathrm{L}$ 205 Mpc]{
    \includegraphics[width=0.48\textwidth]{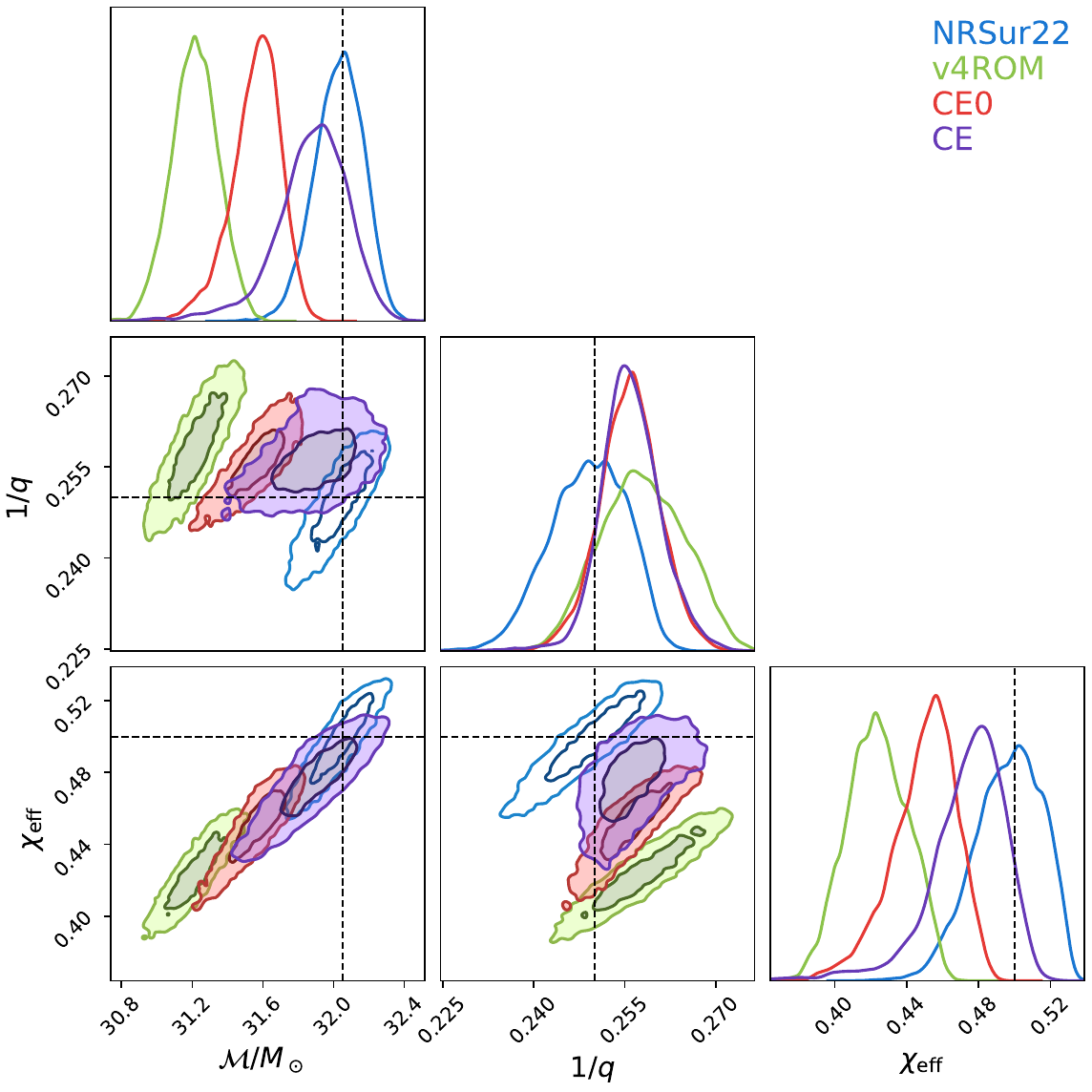}
  }
  \caption{Partial corner plots illustrating the posterior distributions for
  chirp mass  $\mathcal{M}$, mass ratio  $q$, and effective spin $\chieff$,
  derived from Bayesian \ac{PE} using four different waveform
  models: \texttt{NRHybSur3dq8}, \texttt{SEOBNRv4}, \texttt{SEOBNRv4CE}, and \texttt{SEOBNRv4CE0}. 
  The left panel shows results for a non-spinning system, while the right panel displays results 
  for a spinning system with $\chi_1 = \chi_2 = 0.5$ and mass ratio $q=4$. The diagonal panels 
  present the marginalized one-dimensional posterior 
  distributions for each parameter, while the off-diagonal panels depict the two-dimensional 
  joint posterior distributions between pairs of parameters. The contours in the two-dimensional 
  plots represent the 50\% and 90\% credible regions. The true injected values of the parameters 
  are indicated by the dashed black lines.}
  \label{fig:corner_q4}
\end{figure*}

From the corner plot shown in Fig.~\ref{fig:corner_q4}, we can make several key observations.
First, in both the non-spinning and spinning cases at mass-ratio $q=4$, the \texttt{NRHybSur3dq8}
model accurately estimates the posterior, as it is the model used for the signal. In the left panel,
which shows the non-spinning system, the bias in the \texttt{SEOBNRv4} model is small, especially
for chirp mass and mass-ratio. However, in the right panel (spinning system with
$\chi_\mathrm{eff} = 0.5$), significant bias is observed across all three parameters shown in
the plots, with almost no overlap in chirp mass and $\chi_{\text{eff}}$. 
When using the \texttt{SEOBNRv4CE} model, 
we observe bias in all recovered parameters ($\mathcal{M}$, $q$, and $\chieff$) in the non-spinnning case,
yet much of the bias apparent for \texttt{SEOBNRv4} in the spinning system is mitigated by \texttt{SEOBNRv4CE}. \texttt{SEOBNRv4} is expected to be less accurate for spinning systems.
Lastly, we notice strong correlations amongst the BBH parameters, which may pose 
challenges for the \texttt{SEOBNRv4CE} model in effectively exploring the full parameter space. 
These correlations could explain some limitations in parameter recovery using \texttt{SEOBNRv4CE} model.
We also point out that \texttt{SEOBNRv4CE0} and \texttt{SEOBNRv4} lead to different biases.
For the non-spinning configuration in the left panel of Fig.~\ref{fig:corner_q4} \texttt{SEOBNRv4CE}
and \texttt{SEOBNRv4CE0} show similarly large bias, much worse than the bias found for
\texttt{SEOBNRv4}. However, for the spinning configuration shown in the right panel
\texttt{SEOBNRv4CE} mitigates bias found for \texttt{SEOBNRv4CE0} and
\texttt{SEOBNRv4CE0} performs better than \texttt{SEOBNRv4}.

\subsubsection{Normalized bias over parameter space}

%--------------------------------------------------------------------------
%                        Normalized Bias and Network SNR - 410 Mpc
%--------------------------------------------------------------------------
\begin{figure*}[!htbp]
  \centering
  \subfloat[Normalized bias]{
    \hspace{-15pt}
    \includegraphics[width=0.7\textwidth]{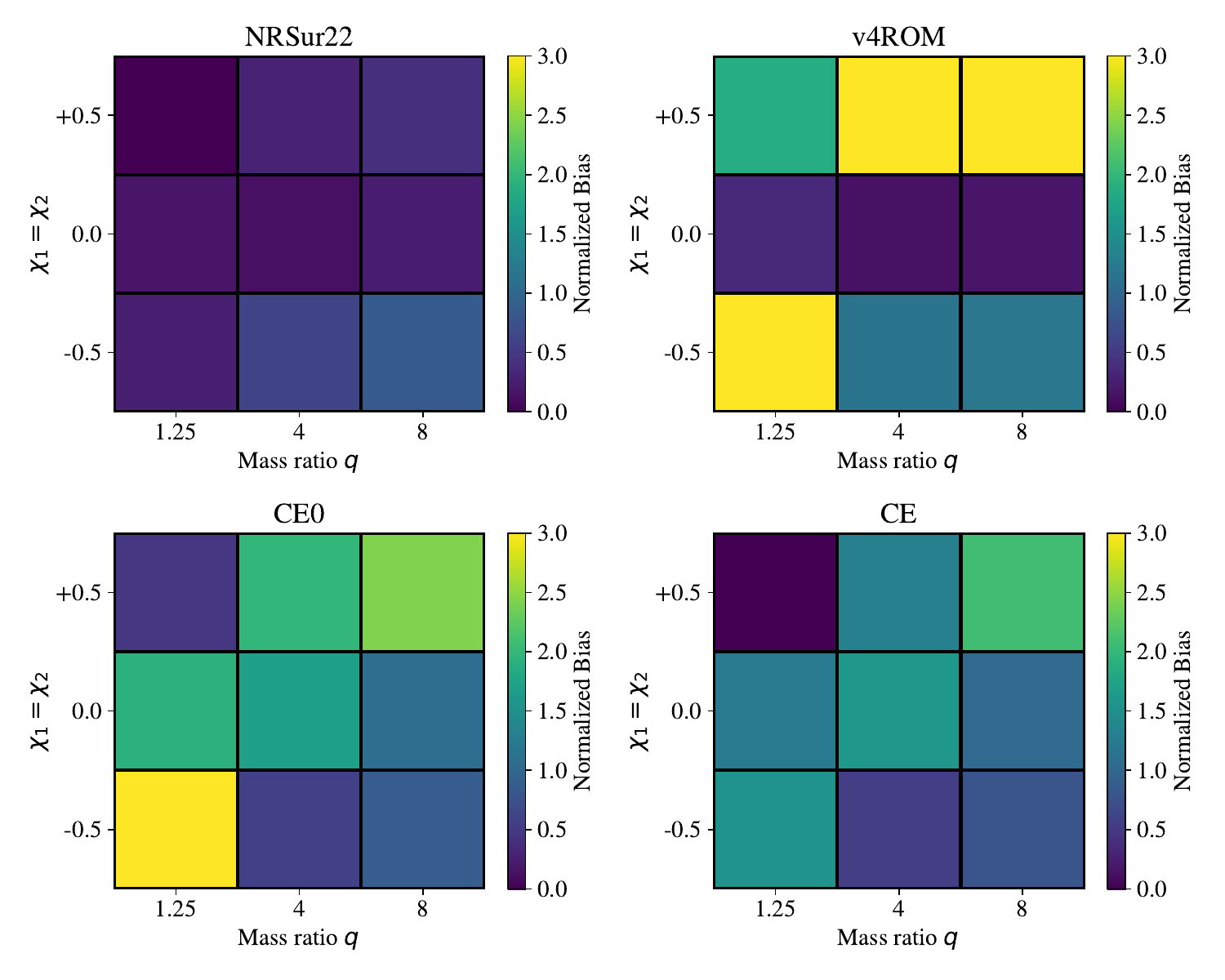}
    \label{fig:normalized_bias1_410Mpc}
  }
  \subfloat[Network SNR]{
    \hspace{-10pt}
    \raisebox{80pt}{\includegraphics[width=0.35\textwidth]{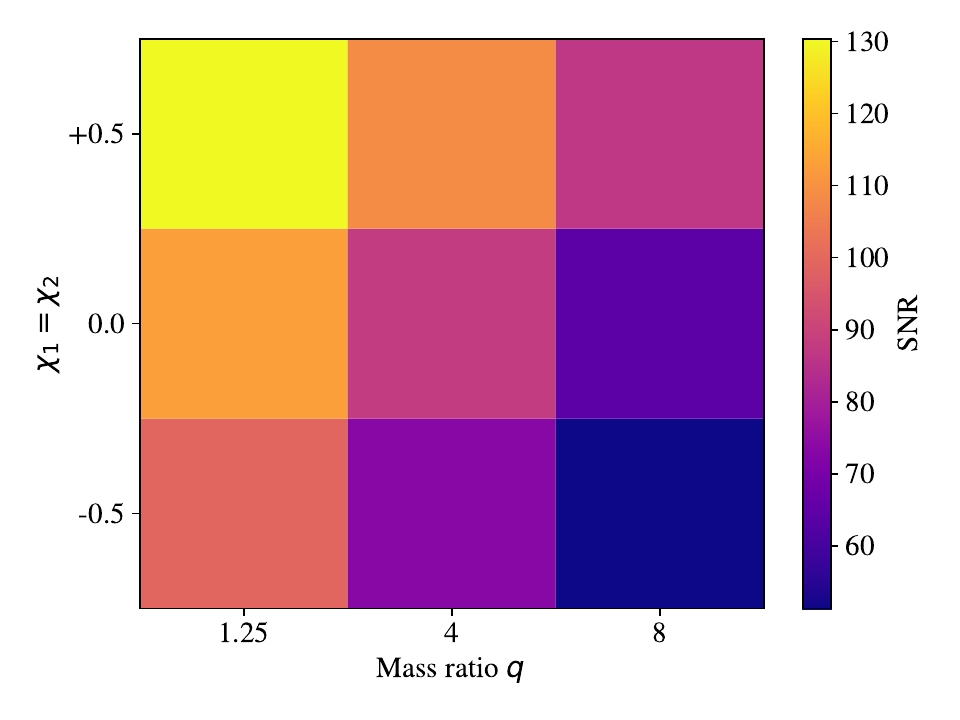}}
    \label{fig:snr_410Mpc}
  }
  \caption{
    Left figure:  Normalized biases for chirp mass over a grid of \texttt{NRHybSure3q8} signals with
    varying mass ratio $q$ and aligned spins $\chi_1 = \chi_2$ at a luminosity distance of 410 Mpc.
    The signals are recovered with four waveform models \emph{top left:} \texttt{NRHybSure3q8},
    \emph{top right:} \texttt{SEOBNRv4}, \emph{bottom left:} \texttt{SEOBNRv4CE0}, and 
    \emph{bottom right:} \texttt{SEOBNRv4CE}. Right figure shows the network SNR of the injected
    signals for the given luminosity distance.
  }
\end{figure*}

%--------------------------------------------------------------------------
%                        Normalized Bias - 820 Mpc
%--------------------------------------------------------------------------
\begin{figure*}[!htbp]
  \centering
  \subfloat[Normalized bias]{
    \hspace{-15pt}
    \includegraphics[width=0.7\textwidth]{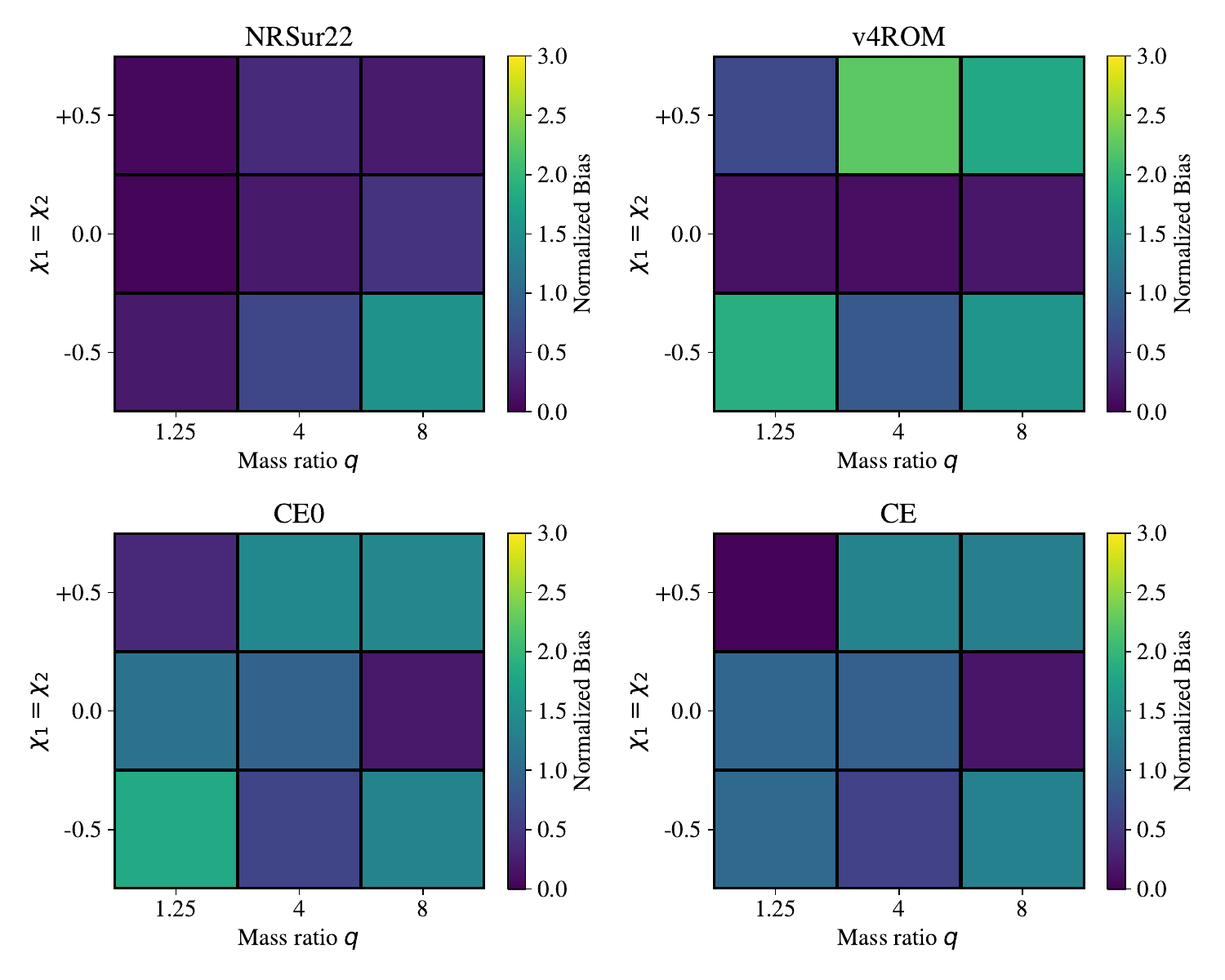}
    \label{fig:normalized_bias1_820Mpc}
  }
  \subfloat[Network SNR]{
    \hspace{-10pt}
    \raisebox{80pt}{\includegraphics[width=0.35\textwidth]{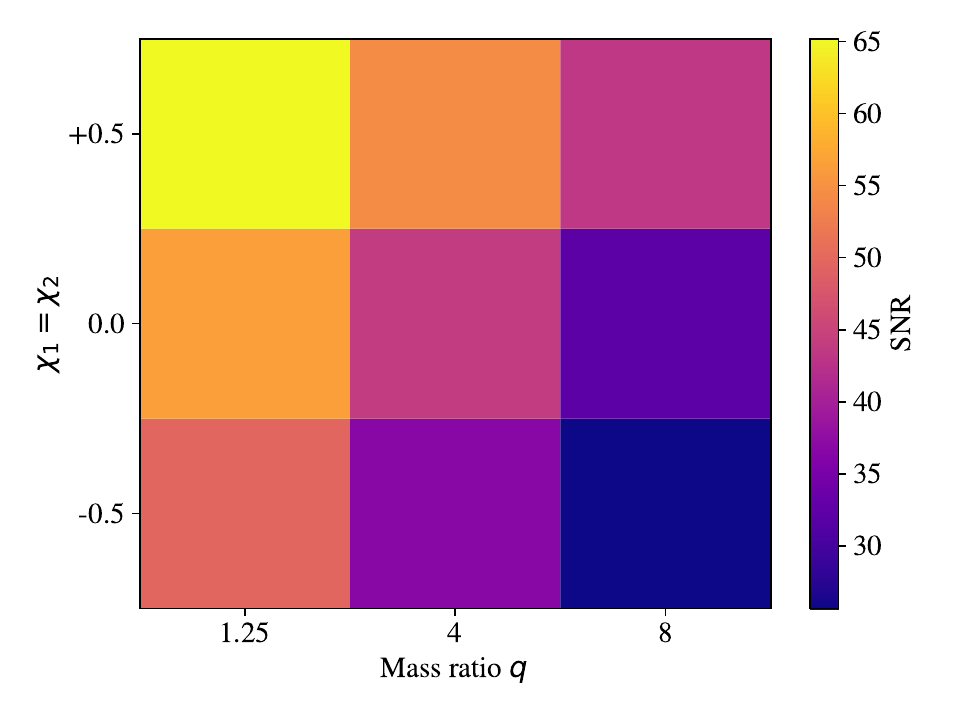}}
    \label{fig:snr_820Mpc}
  }
  \caption{
    Normalized biases for chirp mass as shown in Fig.~\ref{fig:normalized_bias1_410Mpc}
    except for a farther distance $d_\mathrm{L} = 820\,\mathrm{Mpc}$, which corresponds to lower SNRs. 
    For ease of comparison with Fig.~\ref{fig:normalized_bias1_410Mpc} we have kept the same
    limits on the normalized bias scale and colormap.
    Corresponding network SNR is given in the right figure.
  }
\end{figure*}

In Fig.~\ref{fig:normalized_bias1_410Mpc} we show normalized bias in chirp mass for signals varying 
over a $3$ by $3$ grid of mass-ratios and spins for a luminosity distance of $d_\mathrm{L} = 410$  Mpc.
As expected, the bias is minimal for recovery with the \texttt{NRHybSur3dq8} model. 
The bias tends to increase for the \texttt{SEOBNRv4} model, especially for spinning configurations 
$\chi_1 = \chi_2 = \pm 0.5$. Much of this bias is mitigated with the \texttt{SEOBNRv4CE} model.
However, we observe an overall increase in bias for non-spinning systems with the uncertainty 
model, compared to the \texttt{SEOBNRv4} model. This may be due to the \texttt{SEOBNRv4} model 
being better calibrated against \ac{NR} simulations near zero spins~\cite{Bohe:2016gbl} and the
\texttt{SEOBNRv4CE} being less well calibrated there, as is also indicated by increased bias
seen for \texttt{SEOBNRv4CE0} for non-spinning binaries. Here, \texttt{SEOBNRv4CE0} tends to perform
better than \texttt{SEOBNRv4} for positive aligned spins $\chi_1 = \chi_2 = + 0.5$.

Because of our choice of constant distance for all signals shown in Fig.~\ref{fig:normalized_bias1_410Mpc}
the SNR varies with mass ratio and spin. The network SNR corresponding to each configuration is
shown in the Fig. \ref{fig:snr_410Mpc}. The comparable mass binary with positive aligned spin 
has highest SNR $\sim 130$.

Similarly, we plot the normalized bias for a set of corresponding configurations with luminosity 
distance $d_\mathrm{L} = 820$ Mpc in Fig~\ref{fig:normalized_bias1_820Mpc}, which constitutes a 
lower SNR version of the systems shown in Fig.~\ref{fig:normalized_bias1_410Mpc}. As expected, 
at lower SNRs, configurations are more affected by statistical errors, leading to broader posteriors. 
Consequently, the overall bias in chirp mass $\mathcal{M}$ is reduced across the parameter space.

\subsubsection{Idealized scaling}
\label{ssub:idealized_scaling}

\begin{figure}[!h]
  \includegraphics[width=0.45\textwidth]{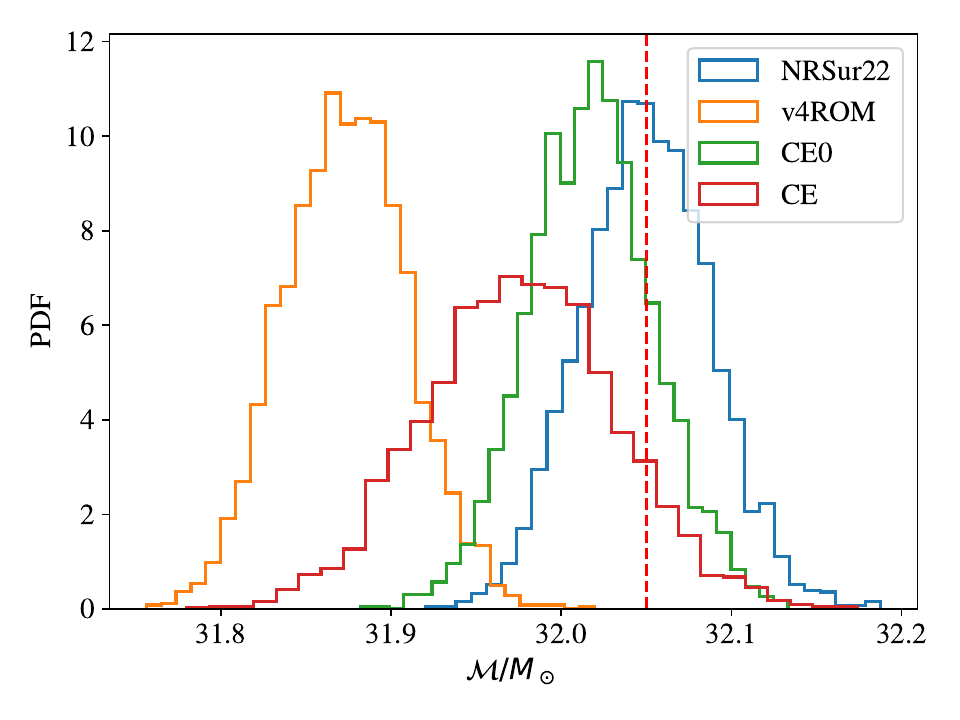}
  \caption{
    Chirp mass posterior of GW150914-like \texttt{NRHybSur3dq8} signal recovered with
    waveform models \texttt{NRHybSur3dq8}, \texttt{SEOBNRv4}, \texttt{SEOBNRv4CE0}, and
    \texttt{SEOBNRv4CE}.
    The network SNR corresponding to this event is $111.5$. 
  }
  \label{fig:gw150914_chirp_mass}
\end{figure}

\begin{figure}[!h]
  \includegraphics[width=0.45\textwidth]{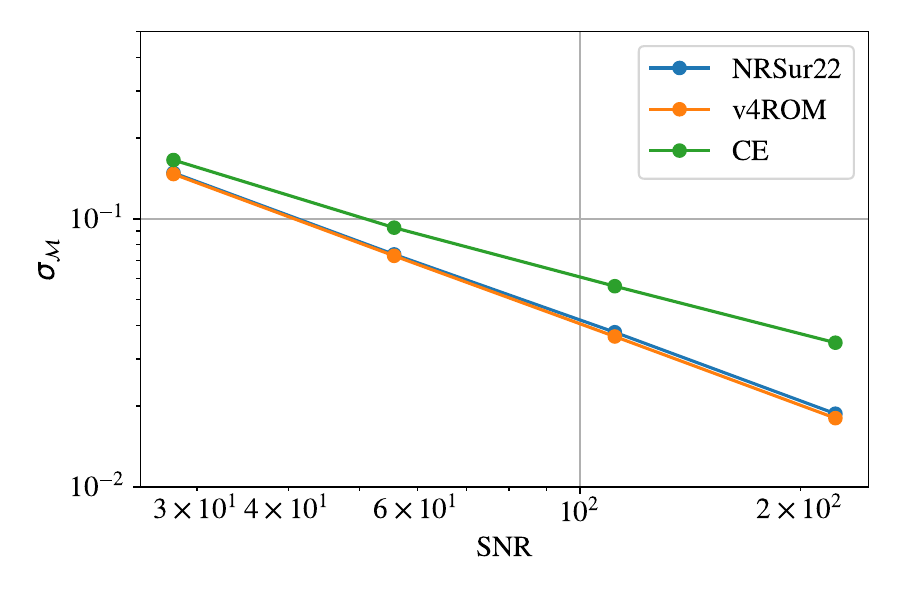}
  \caption{
    Standard deviation of chirp mass posterior as a function of SNR for the GW150914-like signal
    shown in Fig.~\ref{fig:gw150914_chirp_mass}.
    We find slopes $\alpha = -0.99$ for \texttt{NRHybSur3dq8}, $\alpha = -1.00$ for \texttt{SEOBNRv4},
    and $\alpha = -0.75$ for \texttt{SEOBNRv4CE}.
  }
  \label{fig:gw150914_snr_vs_sigma}
\end{figure}

Finally, we seek to test our uncertainty model in a more simplistic situation. We select a
\emph{GW150914}-like signal using the \texttt{NRHybSur3dq8} model. Accordingly we choose
mass ratio $q = 1.228$, and aligned spins $\chi_1 = -0.32$, $\chi_2 = -0.57$, and luminosity distance $d\mathrm{L} = 410$~Mpc \cite{LIGOScientific:2016ebw, LIGOScientific:2016aoc}.
The remaining parameters are set as for the other configurations discussed in Sec.~\ref{sub:PE_setup}.
For this analysis, we only sample in chirp mass $\mathcal{M}$, coalescence time $t_c$, and relative phase $\phi_c$. All other parameters are pinned to the true parameter value of the signal.
Fig.~\ref{fig:gw150914_chirp_mass} shows the chirp mass posteriors. We witness significant bias
for \texttt{SEOBNRv4} model with no support at the true value.
\texttt{SEOBNRv4CE} mitigates some of the bias in the chirp mass. 
Here, \texttt{SEOBNRv4CE0} performs very well showing only a small amount of bias.
In Fig.~\ref{fig:gw150914_snr_vs_sigma} we show the standard deviation of the chirp mass $\sigma_{\mathcal{M}}$ as a function of \ac{SNR}. In this idealistic situation both
\texttt{NRHybSur3dq8} and \texttt{SEOBNRv4} have a close to ideal power exponent $\alpha \sim -1$.
For \texttt{SEOBNRv4CE} we obtain $\alpha = -0.75$, which again corresponds to wider posterior
distributions compared to models which do not include waveform uncertainty for all values of \ac{SNR}.

%--------------------------------------------------------------------------
%             GW150914 - chirp mass posterior (pin standard)
%--------------------------------------------------------------------------

%--------------------------------------------------------------------------
%             GW150914 - SNR vs Sigma (pin standard)
%--------------------------------------------------------------------------

%==========================================================================
\section{Conclusion}
\label{sec:conclusion}
 
In this paper, we develop a novel probabilistic waveform model, \texttt{SEOBNRv4CE}, as an alternative
to the deterministic \texttt{SEOBNRv4} model. As discussed in Sec.~\ref{ssec:uncertainty_model}
\texttt{SEOBNRv4CE} uses a multivariate normal approximation, Eq.~\eqref{eq:mvn_ansatz}, of amplitude
and phase deviations of \ac{GW} data from the calibrated \texttt{SEOBNRv4} model and the uncalibrated
\texttt{SEOBNRv4} model evaluated at draws from the calibration posterior samples $p(\theta | \lambda_i)$.
We use Gaussian Process Regression (GPR) to interpolate the means and Cholesky decomposition of
covariance matrices of these deviations. The resulting model
$\tilde h_\mathrm{CE}(\lambda, \vec\epsilon; f)$ consisting of Eqs.\eqref{eq:seobnrv4ce_model}, \eqref{eq:modeled_differences}, and \eqref{eq:epsilon_normal}
can be efficiently computed in the Fourier domain. We predict the amplitude and phase deviations at ten
frequency nodes and therefore \texttt{SEOBNRv4CE} depends on twenty parameters $\vec\epsilon$ which
capture \ac{NR} calibration uncertainty in the \texttt{SEOBNRv4} model.

Unlike \citet{Moore:2014pda} who directly model waveform deviations $\delta h$ we cannot analytically marginalize over waveform uncertainty because of our amplitude phase parametrization, even though we are
also using \ac{GPR}. We perform a Bayesian inference study where we marginalize over all uncertainty
parameters. By doing so, we are able to incorporate waveform systematics arising from the calibration
process in the \texttt{SEOBNRv4} model.

We evaluate the effectiveness of the \texttt{SEOBNRv4CE} model by performing a detailed \ac{PE}
study using mock \ac{GW} signals described in Sec.~\ref{sec:PE_campaign}. We analyze \texttt{NRHybSur3dq8}
signals in zero noise for variety of \ac{BBH} configurations over the binary parameter space and recover
the posterior using four template models -- \texttt{NRHybSur3dq8}, \texttt{SEOBNRv4}, \texttt{SEOBNRv4CE},
and \texttt{SEOBNRv4CE0}.
We find that a \ac{PE} analysis using \texttt{SEOBNRv4CE} takes only a factor of two longer than
an analysis with \texttt{SEOBNRv4}, even though the uncertainty model needs to (i) evaluate the \ac{GPR}
of amplitude and phase deviations, compute a complex exponential, and multiply the corrections with the 
base model, and (ii) sample in twenty additional uncertainty parameters.
Our results demonstrate that, for most cases, the uncertainty model reduces bias compared to the
standard \texttt{SEOBNRv4} model in parameter recovery. Such bias is expected for high SNR signals
and when waveform systematics dominate over statistical uncertainties. In particular, we observe a broadening of \texttt{SEOBNRv4CE} posteriors compared to those of other models (see
Figs.~\ref{fig:snr_vs_sigma} and \ref{fig:gw150914_snr_vs_sigma}), together with a reduction in bias
for most configurations (see Figs~\ref{fig:marginal_chirp_mass}, \ref{fig:corner_q4}, and
\ref{fig:gw150914_chirp_mass}).

Our investigation of normalized biases over parameter space (see Fig.~\ref{fig:normalized_bias1_410Mpc})
reveals that systematic biases for \texttt{SEOBNRv4} arise predominantly for spinning configurations,
for which \texttt{SEOBNRv4} is less well calibrated against \ac{NR} than for non-spinning configurations. This is especially true for unequal mass configurations with moderately high positive aligned spin, but
we also encountered biases for a near equal mass configuration with negative aligned spins. Some of this
effect may be due to the lower density of \ac{NR} simulations available for calibration, and some
due to the challenge of accurately modeling the more complicated physics of spinning binaries.

The uncertainty model proved to be effective in mitigating biases for spinning configurations -- 
reducing normalized biases by a factor of $\sim 1.5$ or more;
however, for non-spinning binaries, where \texttt{SEOBNRv4} is already well calibrated, \ac{PE} with \texttt{SEOBNRv4CE} model fell short of mitigating biases. This limitation may stem form the 
simplifying assumption that the metric entering the GPR covariance function,
Eq.~\eqref{eq:gpr-covariance-function} in Sec.~\ref{sub:gaussian_process_regression} is diagonal,
containing learnable length scale hyper-parameters for mass ratio and aligned spins.
Realistically, one should take into account correlations between intrinsic BBH parameters in the 
choice of \ac{GPR} hyper-parameters. We may explore this possibility in a future work.
%

% \RB{gw150914-like event analysis with pinned standard parameters}
We demonstrated that chirp mass posteriors for \texttt{SEOBNRv4CE} are wider in terms of the standard
deviation than those of models which do not include waveform uncertainty. In an idealized setting
where most waveform parameters are pinned and we only sample in chirp mass, time and phase, we found
that the standard deviations for \texttt{SEOBNRv4} or \texttt{NRHybSur3dq8} decrease inversely with SNR,
in agreement with the behavior of a Gaussian posterior, whereas the standard deviation for
\texttt{SEOBNRv4CE} decreases more slowly, at a power-law rate of $-0.75$, compared to the ideal exponent
of $-1$, as also predicted by \citet{Moore:2014pda} (see the increased variance in
Eq.~\eqref{eq:marg_likelihood}).

% Summarize literature
\citet{Moore:2014pda} proposed a likelihood function which marginalizes analytically over waveform uncertainty using a Gaussian process prior on the waveform difference $\delta h$. Their analysis was restricted to sampling the chirp mass and they assumed the model errors to be frequency independent. Our uncertainty model takes into account correlations in frequency for amplitude and phase at a set of frequency nodes.
\citet{Edelman:2020aqj} instead explored the implications of unmodeled physics in \ac{GW} observations.
They augment an existing waveform model with unknown amplitude and phase deviations in the Fourier domain and infer posterior distributions for these deviations to constrain potential deviations from known physics accurately, instead of marginalizing over them.
\citet{Read:2023hkv} introduces a frequency-dependent characterization of waveform 
error in amplitude and phase for binary neutron star signals and demonstrates how these differences can be marginalized over during the inference using technique similar to those applied in detector calibration. 
Lastly, \citet{Owen:2023mid} use Bayesian inference to assess the inaccuracies in theoretical waveform models due to missing higher-order \ac{PN} terms. By treating these missing terms as unknown parameters and marginalizing over them with uninformed priors, the authors show improved parameter estimates, at the cost of higher statistical error.

While techniques for modeling and incorporating waveform uncertainty into \ac{PE} have been explored
in a few studies, they are not yet used as part of routine analyses carried out by the LVK collaboration.
Clearly, more work needs to be done on the waveform side to construct waveform uncertainty models
for more complete waveform models, including important physical effects such as precession, and
eccentricity, and including higher order modes. However, such techniques also need to be efficient
and reliable before they will find adoption in large scale data analysis. Given that progress in 
directly improving the accuracy of waveform models to the levels required of third generation detectors
will be very challenging, constrained by practical modeling assumptions and, most importantly, the availability of a dense set of high accuracy \ac{NR} simulations covering
binary parameter space, marginalization over waveform uncertainty should be a promising path to follow
to enhance the fidelity of parameter estimates.

%==========================================================================
\begin{acknowledgments}
We wish to thank Christopher Moore, Vivien Raymond, Roberto Cotesta, Sergei Ossokine, and Alessandra Buonanno
for contributions during the early part of this project.
We would like to thank Jonathan Gair and Lorenzo Pompili, and Carl Haster for helpful discussions.
The authors acknowledge support from the URI Center for Computational Research. In particular, 
the computations were performed on the UMass-URI \texttt{UNITY} high-performance computing cluster 
as well as the \texttt{Hypatia} computer cluster at the Max Planck Institute for Gravitational 
Physics in Potsdam-Golm, Germany.
% LIGO
This material is based upon work supported by NSF's LIGO Laboratory which is a major facility fully funded by the NSF.
\end{acknowledgments}

%%%%%%%%%%%%%%%%%%%%%%%%%%%%%%%%%%%%%%%%%%%%%%%%%%%%%%%%%%%%%%%%%%%%%%%%%%%%%%%
\section*{References}
%%%%%%%%%%%%%%%%%%%%%%%%%%%%%%%%%%%%%%%%%%%%%%%%%%%%%%%%%%%%%%%%%%%%%%%%%%%%%%%
\bibliography{References}

\end{document}